\documentstyle[prl,aps,epsfig,multicol,amssymb,amsfonts]{revtex}
\tighten
\renewcommand{\narrowtext} 
{\begin{multicols}{2}\global\columnwidth20.5pc} 
\renewcommand{\widetext}
{\end{multicols}\global\columnwidth42.5pc} 
\newcommand{\be}{\begin{equation}}
\newcommand{\ee}{\end{equation}}
\newcommand{\bea}{\begin{eqnarray}}
\newcommand{\eea}{\end{eqnarray}}
\newcommand{\br}{{\bf r}}

\begin{document} 
\draft 
\title{Wave function statistics and multifractality at the spin
quantum Hall transition} 
\author{A.~D.~Mirlin$^{1,2,*}$, F.~Evers$^1$, and A.~Mildenberger$^1$}  
\address{$^1$Institut
f\"ur Nanotechnologie, Forschungszentrum Karlsruhe, 76021 Karlsruhe,
Germany}
\address{$^2$Institut f\"ur Theorie der Kondensierten Materie,
Universit\"at Karlsruhe, 76128 Karlsruhe, Germany}
\date{\today}
\maketitle
\begin{abstract}
The statistical properties of wave functions 
at the critical point of the spin quantum Hall transition are
studied. The main emphasis is put onto determination of the spectrum
of multifractal exponents $\Delta_q$ governing the scaling of moments
$\langle|\psi|^{2q}\rangle\sim L^{-qd-\Delta_q}$ with the system size
$L$ and the spatial decay of wave function correlations. Two- and
three-point correlation functions are calculated analytically by means
of mapping onto the classical percolation, yielding the values
$\Delta_2=-1/4$ and $\Delta_3=-3/4$. The multifractality spectrum
obtained from numerical simulations is given with a good 
accuracy by the parabolic approximation $\Delta_q\simeq q(1-q)/8$
but shows detectable deviations. We also study statistics of the
two-point conductance $g$, in particular, the spectrum of exponents
$X_q$ characterizing the scaling of the moments $\langle
g^q\rangle$. Relations between the 
spectra of critical exponents of wave functions ($\Delta_q$),
conductances ($X_q$), and Green functions at the localization transition
with a critical density of states are discussed. 
\end{abstract}

\narrowtext

\section{Introduction}
\label{s1}

In the framework of the random matrix theory pioneered by Wigner
\cite{wigner} and Dyson \cite{dyson}, the statistical properties of
spectra of complex systems are described by random matrix ensembles. 
Within the Dyson's classification, three symmetry classes are
distinguished (orthogonal, symplectic, and unitary), depending on
whether the system is invariant under the time-reversal transformation
and on its spin. It has been understood that this classification is
very general and applies to a great variety of physically distinct
systems (see \cite{guhr} for a recent review).

While the Dyson's classification is complete for the bulk of the
spectrum, more symmetry classes may arise in the vicinity of a special
point on the energy axis. Such non-standard symmetry classes have
attracted a considerable research attention during the last
decade. One group of them is formed by three chiral ensembles
\cite{verbaarschot} describing the spectrum of a massless Dirac
operator near zero energy. The same symmetry is shared by
tight-binding models with purely off-diagonal disorder at the band
center \cite{gade}. More recently, four more symmetry classes were
identified \cite{altland97}, which characterize a dirty superconductor
or a mesoscopic 
superconductor-normal metal system. The Hamiltonian matrix has in this
case an additional block structure in the particle-hole space induced
by the form of mean-field Bogoliubov-de Gennes equations for a
superconductor. It was argued \cite{zirnbauer} that the extended
classification scheme including 10 classes (three Wigner-Dyson, three
chiral, and four Bogoliubov-de Gennes) is complete.

The classification of random matrix ensembles can be equally well
applied to disordered electronic systems. In particular,
two-dimensional systems of 
non-standard classes are of large interest, in view of
their relevance to high-$T_c$ superconductors, which have an
unconventional ($d$-wave) symmetry of the order parameter and
therefore possess low-energy quasiparticle excitations. In this
paper, we will consider a system of class C, which is the
Bogoliubov-de Gennes class with broken time-reversal but preserved
spin rotation invariance. The corresponding Hamiltonian satisfies the
symmetry $H^*=-\sigma_yH\sigma_y$ (with $\sigma_y$ the Pauli matrix in
the particle-hole space) and has the block structure
\be
\label{e0}
H=\left(\begin{array}{ll} h           & \Delta \\
                         \Delta^*    & -h^T
\end{array}\right),
 \ee
where $h=h^\dagger$ and $\Delta=\Delta^T$. 

Similarly to the conventional Wigner-Dyson unitary class, a
two-dimensional system of class C undergoes a transition between the
phases with different quantized values of the Hall conductivity
\cite{kagalovsky99,senthil99,gruzberg99}. More 
precisely, since the quasiparticle charge is not conserved in a
superconductor, one is led to consider the spin conductivity
determining the spin current as a response to the gradient of the
Zeeman magnetic field. The quantization of the Hall component of the
spin conductivity tensor was named the spin quantum Hall (SQH) effect. 
It was shown \cite{senthil99} that the SQH effect can be realized in
superconductors with $d_{x^2-y^2}+id_{xy}$ pairing symmetry
explored in recent literature \cite{d-plus-id}.

While the SQH transition shares many common features with its normal
counterpart, it is qualitatively different as concerns the behavior of
the density of states (DOS) at criticality: while the DOS is
uncritical for the conventional quantum Hall (QH) transition, it
vanishes at the SQH critical point. 
A network model describing the SQH transition was
constructed in \cite{kagalovsky99}, and critical exponents for the scaling
of the localization length were determined numerically. In
\cite{senthil99} a mapping onto a supersymmetric spin chain was
performed, providing an alternative method for the numerical study of the
critical behavior. 
Remarkably, some exact analytical results for this problem have been
obtained by mapping onto the classical percolation
\cite{gruzberg99,cardy00,beamond02}. Specifically, it was found that
the DOS scales as $\rho(\epsilon)\sim \epsilon^{1/7}$,
while the average product of the retarded and advanced Green functions
$\Pi({\bf r},{\bf r}')=\langle G_R(\br,\br')G_A(\br',\br)\rangle$
(referred to as the diffusion propagator, or the diffuson)
and the average two-point conductance 
$\langle g({\bf r},{\bf r}')\rangle$ at $\epsilon=0$  
fall off as $|{\bf r}-{\bf r}'|^{-1/2}$.

It is known that critical wave functions at the conventional QH
transition have multifractal nature \cite{huckestein95,janssen98}. 
Recently, there has been a growth 
of activity in the direction of quantitative characterization of the
corresponding spectrum of fractal dimensions
\cite{janssen99,zirnbauer99,bhaseen00,klesse01,evers01}. Zirnbauer
\cite{zirnbauer99}  and Bhaseen {\it et al.} \cite{bhaseen00} proposed
a certain supersymmetric $\sigma$-model with a Wess-Zumino-Novikov-Witten
term (in two slightly different versions) as a candidate for the
conformal field theory of the QH critical point. The theory implies
an exactly parabolic form of the multifractality spectrum. This was
confirmed by a thorough numerical study of the wave function
statistics at the QH transition \cite{evers01}. 

The aim of this paper is to study the wave function statistics at the
SQH critical point. We will demonstrate that the exponents $\Delta_2$
and $\Delta_3$  governing the scaling of the second and third moments
of the wave function intensity (see Sec.~\ref{s2} for the formal
definition) can be calculated exactly by analytical means. Quite
surprisingly, we find that the index $\eta=-\Delta_2$ characterizing
the spatial decay of the wave function correlations is equal to $1/4$,
in contrast to the $r^{-1/2}$ decay of the diffusion propagator. This
leads us to a general analysis of relations between different critical
exponents characterizing the wave function statistics in the
qualitatively new situation of the localization transition with a
critical DOS. We complement our analytical results by numerical
simulations, which allow us, in particular, to investigate whether the
multifractality spectrum of the SQH critical point is exactly
parabolic or not. The answer to this question, as well as the exact
values of $\Delta_2$ and $\Delta_3$ we have found, is of central
importance for identification of conformal theory of the SQH
transition, which is the issue of a considerable research interest
at present \cite{fendley00,bernard01a,bernard01b}. Some of our results
were reported in a brief form in \cite{sqhe-letter}.

The article is organized as follows. In Sec.~\ref{s2} we remind the
reader of some basic concepts related to the multifractality of
critical wave functions. In Sec.~\ref{s3} we describe the network model of
class C and use it to calculate numerically the DOS at the critical
point of the SQH transition. In Sec.~\ref{s4} we present an analytical
calculation which involves a mapping onto the percolation theory and
allows us to calculate the averages of products of two and three
Green's functions and thus the exponents $\Delta_2$ and
$\Delta_3$. Section~\ref{s5} is devoted to a numerical evaluation of
the full multifractal spectrum $\Delta_q$. This allows us not only to
check the analytical results of Sec.~\ref{s4} but also to investigate
whether the spectrum is exactly parabolic (as for the conventional QH
critical point) or not. In Sec.~\ref{s6} we present a numerical study
of statistical properties of the two-point conductance. We further
include a scaling analysis of the relation between the multifractal
spectra of the two-point conductance and of the wave functions at a
critical point with a vanishing DOS. These analytical arguments
clarify the connection between the numerical findings of Sec.~\ref{s6}
and the results of Sec.~\ref{s4}, \ref{s5} on the wave function
multifractality. Finally, Sec.~\ref{s7} contains a summary of our
results and a brief discussion of some remaining open problems. 

\section{Wave function multifractality in systems with non-critical DOS}
\label{s2}

Multifractality of wave functions $\psi({\bf r})$ is known to
be a hallmark of the localization transition.
It has been extensively studied in the context of conventional Anderson and
quantum Hall (QH) transitions with {\it non-critical} DOS (see
\cite{huckestein95,janssen98,adm-review} and references therein), and
we remind the 
reader of some basic results. Multifractality is characterized by a set
of exponents 
\be
\label{e1a}
\tau_q\equiv d(q-1)+\Delta_q
\ee
($d$ is the spatial
dimensionality) describing the scaling of the
moments of $|\psi^2({\bf r})|$ with the system size $L$, 
\be
\label{e1b}
\langle|\psi({\bf r})|^{2q}\rangle \sim L^{-d-\tau_q}.
\ee  
Anomalous dimensions $\Delta_q$ distinguish a
critical point from the metallic phase and determine the scale dependence
of wave function correlations. Among them, $\Delta_2\equiv -\eta$ plays
the most prominent role, governing the spatial correlations of
the ``intensity'' $|\psi|^2$,
\be
L^{2d} \langle |\psi^2({\bf r})\psi^2({\bf r}')|\rangle 
\sim (|{\bf r} - {\bf r}'|/L)^{-\eta}.
\label{e1}
\ee
Equation (\ref{e1}) can be obtained from (\ref{e1b}) by using the fact
that the wave function amplitudes become essentially uncorrelated at
$|\br-\br'|\sim L$. Scaling behavior of higher order spatial
correlations, 
$\langle|\psi^{2q_1}({\bf r_1})\psi^{2q_2}({\bf r_2})\ldots
\psi^{2q_n}({\bf r_n})|\rangle$ can be found in a similar way.  
Correlations of two different (but close in energy) eigenfunctions and
the diffusion propagator $\Pi(\br,\br';\omega)=
\langle G^R_{E+\omega}(\br,\br') G^A_E(\br',\br)\rangle$ ($G^{R,A}$
are retarded and advanced Green functions) possess the same scaling properties,
\be
\left.\begin{array}{l}
L^{2d} \langle |\psi_i^2({\bf r})\psi_j^2({\bf r}')|\rangle \\
L^{2d} \langle \psi_i({\bf r})\psi_j^*({\bf r})
\psi_i^*({\bf r}')\psi_j({\bf r}')\rangle \\ 
\rho^{-2}\Pi({\bf r},{\bf r}';\omega) 
\end{array}  \right\}
\sim \left( {|{\bf r} - {\bf r}'| \over L_\omega}\right)^{-\eta},
\label{e2}
\ee
where $\omega=\epsilon_i-\epsilon_j$, 
$L_\omega\sim (\rho\omega)^{-1/d}$, $\rho$ is the density of states, 
and $|{\bf r} - {\bf r}'| < L_\omega$. 
In two dimensions the multifractal spectrum $\Delta_q$ plays a
key role in the identification of the conformal field theory governing
the critical point, which led to growing interest in the
eigenfunction statistics at the QH transition
\cite{janssen99,zirnbauer99,bhaseen00,klesse01,evers01}. 

Applying naively these results to the SQH transition, one would
conclude that the $r^{-1/2}$ scaling of the diffusion propagator found
in \cite{gruzberg99} implies $\eta=1/2$. 
However, we show below that this conclusion is incorrect. This
demonstrates that one should be 
cautious when trying to apply the relations between critical
exponents obtained for systems with a non-critical DOS to those with a
critical one (like the SQH transition), as will be discussed in
Sec.~\ref{s4.3} and Sec.~\ref{s6}.

\section{Network model and the density of states}
\label{s3}

As a model of the SQH system, we use the SU(2) version
\cite{kagalovsky99} of the
Chalker-Coddington network describing the QH transition
\cite{chalker88}. Dynamics 
of the wave function defined on edges of the network is governed by a
unitary evolution operator ${\cal U}$. At each node of the network the
scattering from two incoming into two outgoing links is described by a
matrix
\be
\label{e3a}
S=\left(\begin{array}{ll} \cos\theta & \sin\theta \\
                         -\sin\theta & \cos\theta \end{array}
\right),
\ee
with $\theta=\pi/4$ corresponding to the critical point. Each
realization of the network is characterized by a set of random
$2\times 2$ spin matrices $U_e$ associated with all edges $e$ of the
network. In view of (\ref{e0}), ${\cal U}$ satisfies the symmetry
${\cal U}=\sigma_y{\cal U}^*\sigma_y$, implying that $U_e\in {\rm
SU(2)}$. Diagonalizing ${\cal U}$ 
for a square network of the size $L\times L$ yields $4L^2$ eigenfunctions
$\psi_{i\alpha}(e)$ and eigenvalues $e^{-i\epsilon_i}$, where
$i=1,2,\ldots,4L^2$ and $\alpha=1,2$ is the
spin index. 

\begin{figure}
\centerline{
\includegraphics[width=0.9\columnwidth,clip]{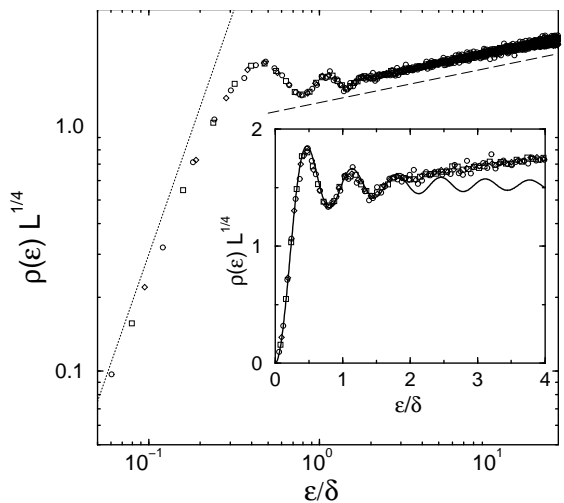}}
\vspace{3mm}
\caption{Scaling plot of the density of states for system sizes
$L=16 (\diamond), 32 (\Box), 96 (\circ)$.
Dashed and dotted lines indicate power laws
(dashed: $\epsilon^{1/7}$, dotted: $\epsilon^{2}$),
$\delta=1/2\pi L^{7/4}$ denotes the level spacing at $\epsilon=0$.
Inset: same data on a linear scale and the result from the
random matrix theory \protect\cite{altland97} (solid curve). 
}
\label{fig1}
\end{figure}

We begin by displaying in Fig.~\ref{fig1} the numerically calculated
DOS $\rho(\epsilon)$ for different system sizes $L$. It is seen that
after a proper rescaling all data collapse onto a single curve. 
Specifically, the energy axis is rescaled to $\epsilon/\delta$, where
$\delta\propto L^{-7/4}$ is the level spacing at $\epsilon=0$. (This
scaling of $\delta$ is related to the critical behavior of DOS
$\rho(\epsilon)\sim\epsilon^{1/7}$ discussed below via the condition
$\rho(\delta)\delta\sim 1/L^2$.) The scale invariance of
$\rho(\epsilon)$ at criticality is reminiscent of the analogous
property of the level statistics at the conventional Anderson or QH
transition (see \cite{adm-review} for a review). 
At $\epsilon\gg\delta$ the critical DOS scales as
$\rho(\epsilon)\sim\epsilon^{1/7}$, in agreement with analytical
predictions  \cite{gruzberg99}. On the other hand, at
$\epsilon\sim\delta$ one observes an oscillatory structure
qualitatively analogous to the behavior found in the random matrix
theory (RMT) for the class C \cite{altland97}. 

Let us note that, strictly
speaking, deviations of DOS from the RMT at $\epsilon\sim\delta$ are
not parametrically small. On the other hand, the numerically found DOS
follows very closely the RMT curve for two oscillation periods. In
other words, the energy scale below which the RMT works (the effective
Thouless energy), while being parametrically of order $\delta$, turns
out to be several times larger. This indicates that there is a
numerical smallness in the problem, and the SQH critical point shows
``close-to-metal'' features (similar to the Anderson transition in
$2+\epsilon$ dimensions with small $\epsilon$). The small value 1/7 of
the DOS exponent is another manifestation of the same fact. 

The states with energies $\epsilon\gg\delta$ are localized with the
localization length $\xi_\epsilon\sim\epsilon^{-4/7}$
\cite{gruzberg99}. For smallest energies $\epsilon\sim\delta$ the
correlation length $\xi_\epsilon$ is of the order of the system
size. In view of their critical nature, these states are expected to be
multifractal, $L^{2q}\langle|\psi_{i\alpha}(e)|^{2q}\rangle\sim
L^{-\Delta_q}$. For $\epsilon\gg\delta$ the multifractality holds
within a region of the extent $\xi_\epsilon$ (outside which the wave
function is exponentially small); hence 
\be
\label{e3b}
L^{2}\langle|\psi_{i\alpha}(e)|^{2q}\rangle\sim
\xi_\epsilon^{-2(q-1)-\Delta_q}\equiv \xi_\epsilon^{-\tau_q}.
\ee
By the same token, spatial correlations are expected to be governed by the
multifractality on scales below $\xi_\epsilon$. In particular, we have
for correlations of two different eigenfunctions with energies
$\epsilon_i,\epsilon_j\sim\epsilon$ 
\be
\label{e3c}
L^{4}\langle|\psi_{i\alpha}(e)\psi_{j\beta}(e')|^2\rangle
\sim(r/\xi_\epsilon)^{\Delta_2}, \qquad r\lesssim \xi_\epsilon
\ee
($r$ is the distance between $e$ and $e'$), and similarly for
higher-order correlators. 
In Sec.~\ref{s4} and \ref{s5} we will demonstrate the multifractality
explicitly and calculate the exponents $\Delta_q$.

\section{Two-and three-point correlation functions: mapping onto
percolation problem}
\label{s4}

In this section, we present an analytical calculation of two-point and
three-point correlation functions, which allows us to find the fractal
dimensions $\Delta_2$ and $\Delta_3$. We use the mapping onto the
classical percolation, 
following the approach of \cite{beamond02}, and demonstrate that it can be
extended on products of two and three Green's functions. 

\subsection{Two-point functions}
\label{s4.1}

Consider a correlation function of two wavefunctions,
\bea
\label{e4}
{\cal D}(e',e;\epsilon_1,\epsilon_2)&=&\langle\sum_{ij\alpha\beta}
\psi_{i\alpha}^*(e)\psi_{j\alpha}(e)\psi_{i\beta}(e')\psi_{j\beta}^*(e')
\nonumber \\
&\times & \delta(\epsilon_1-\epsilon_i)\delta(\epsilon_2-\epsilon_j)\rangle,
\eea
where $e$, $e'$ are two different edges of the network.
Introducing the Green function
$$G(e',e;z)=\langle e'|(1-z{\cal U})^{-1}|e\rangle$$
(which is a $2\times 2$ matrix in the spin space), we express
(\ref{e4}) as
\bea
\label{e5}
&& {\cal D}(e',e;\epsilon_1,\epsilon_2)=(2\pi)^{-2}
\langle{\rm Tr}[G_R(e',e;e^{i\epsilon_1})-G_A(e',e;e^{i\epsilon_1})]
\nonumber\\ 
&& \qquad
\times[G_R(e,e';e^{i\epsilon_2})-G_A(e,e';e^{i\epsilon_2})]\rangle, 
\eea
where $G_{R,A}$ are retarded and advanced Green functions,
$G_{R,A}(e',e;e^{i\epsilon_1})=G(e',e;e^{i(\epsilon_1\pm i0)})$.
We will calculate (\ref{e5}) at zero energy, $\epsilon_{1,2}\to 0$, but
finite level broadening, $\pm i0\to \pm i\gamma$ with $\gamma\ll 1$. 
The scaling behavior
of the correlation function (\ref{e4}) at
$\epsilon_1,\epsilon_2\sim\epsilon$ can then be obtained by
substituting $\epsilon$ for $\gamma$. We thus need to calculate
\bea
\label{e6}
{\cal D}(e',e;\gamma)&=&(2\pi)^{-2}
\langle{\rm Tr}[G(e',e;z)-G(e',e;z^{-1})] \nonumber \\
&\times & [G(e,e';z)-G(e,e';z^{-1})]\rangle,
\eea
with a real $z=e^{-\gamma}<1$.  By the same token, in order to understand
the scaling properties of another correlator of two wave functions,
\bea
\label{e6a}
\tilde{\cal D}(e',e;\epsilon_1,\epsilon_2)&=&\langle\sum_{ij\alpha\beta}
|\psi_{i\alpha}(e)|^2|\psi_{j\beta}(e')|^2
\nonumber \\
&\times & \delta(\epsilon_1-\epsilon_i)\delta(\epsilon_2-\epsilon_j)\rangle,
\eea
we will consider the correlation function
\bea
\label{e7}
\tilde{D}(e',e;\gamma)&=&(2\pi)^{-2}
\langle{\rm Tr}[G(e,e;z)-G(e,e;z^{-1})] \nonumber\\
&\times & {\rm Tr}[G(e',e';z)-G(e',e';z^{-1})]\rangle.
\eea
As discussed in the end of Sec.~\ref{s3}, the scaling behavior of 
(\ref{e6}) and (\ref{e7}) at $r\ll\xi_\gamma$ (where $r$ is the
distance between $e$ and $e'$) is governed by the multifractal properties
of wave functions (specifically, by the exponent $\Delta_2$). The general
strategy of calculation of the correlation functions (\ref{e6}),
(\ref{e7}) is analogous to that used in \cite{beamond02} for the
one-point function ${\rm Tr}G(e,e;z)$. Therefore, we outline only briefly
those steps which generalize directly the calculation in \cite{beamond02},
and  concentrate on qualitatively new aspects. 

The Green functions in  (\ref{e6}), (\ref{e7}) are straightforwardly
represented in the form of a sum over paths
\be
\label{e7a}
G(e,e';z)=\sum_{{\rm paths}\ e'\to e} \ldots \cdot zU_{e_j}s_j \cdot 
zU_{e_{j+1}}s_{j+1}\cdot \ldots,
\ee
where $s_j$ is the corresponding matrix element ($\cos\theta$,
$\sin\theta$, or $-\sin\theta$) of the $S$-matrix between the edges $e_j$
and $e_{j+1}$. Equation (\ref{e7a}) generates a convergent expansion
in powers of $z$ when $|z|<1$; otherwise the identity 
\be
\label{e7b}
G^\dagger(e,e';z)={\bf 1}\cdot\delta_{ee'}-G(e',e;(z^*)^{-1})
\ee
is to be used (in all our calculations $z$ is real, so that $z^*=z$). 
As shown below, each of the double sums over paths obtained by
substituting (\ref{e7a}), 
(\ref{e7b}) in (\ref{e6}) or (\ref{e7}) can be reduced to a single sum over
classical paths (hulls) in  the percolation problem. This remarkable
reduction crucially relies on the following two statements:

\begin{enumerate}
 
 \item Only paths visiting each edge of the network either 0 or 2 times are
 to be taken into account; contributions of all the remaining paths sum up
 to zero, 
 
 \item Using the statement 1, it is easy to see that each node may be
 visited 0, 2, or 4 times. The second statement concerns the nodes visited
 four times. As illustrated in Fig.~\ref{fig2}, there are three possibilities 
 how this may happen; the corresponding contributions have weights (i)
 $\cos^4\theta$, (ii) $\sin^4\theta$, and (iii) $-\sin^2\theta\cos^2\theta$
 from the scattering matrix at this node. The statement is that one can
 equivalently take into account only the contributions (i) and (ii) with
 the weights $\cos^2\theta$ and $\sin^2\theta$, respectively. 

\begin{figure}
\centerline{
\includegraphics[width=0.9\columnwidth,clip]{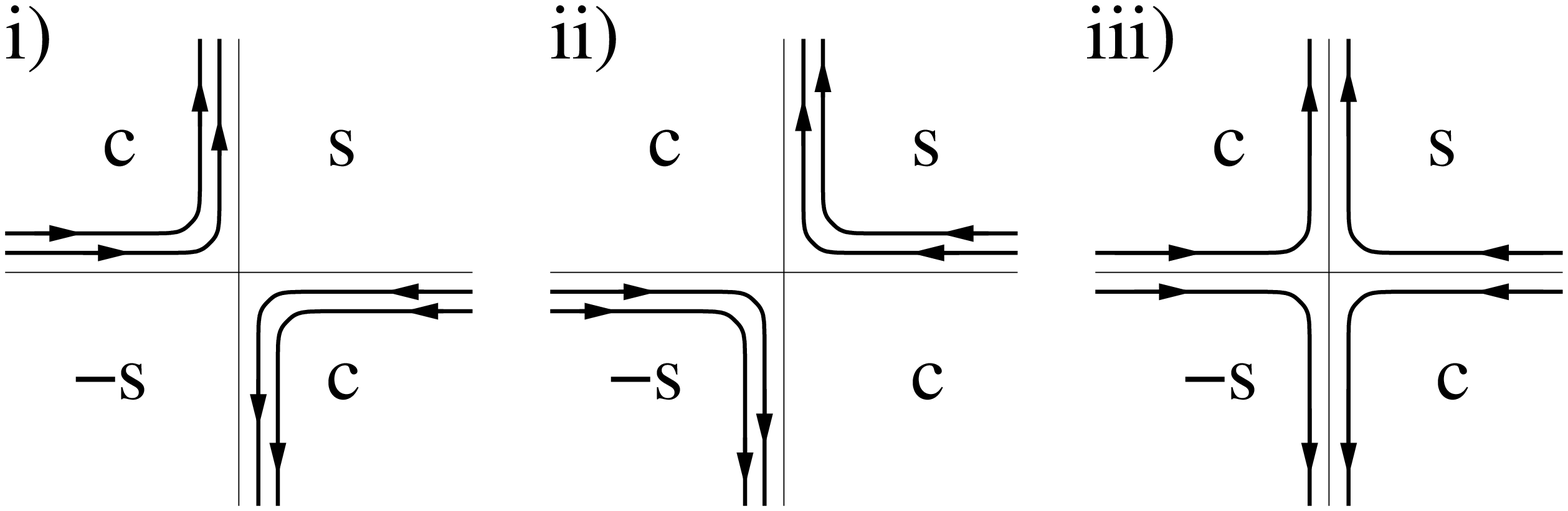}
}
\vspace{3mm}
\caption{Possible configurations of paths passing four times through a
network node. The symbols $c$ and $\pm s$ denote the elements
$\cos^2\theta$, $\pm\sin^2\theta$ of the $S$-matrix at the node. } 
\label{fig2}
\end{figure} 

\end{enumerate}
 
In Ref.~\cite{beamond02} both statements were proven for the case of the
average of a single Green function $\langle G(e,e;z)\rangle$. We show
below  that they are valid for all the two-point functions entering
(\ref{e6}), (\ref{e7}), as well as for averaged products of three
Green's functions 
(considered in Sec.~\ref{s4.2}). Let us emphasize that such a
generalization is far from trivial. This point is well illustrated by the
fact that products of four (or more) Green functions determining the
exponents $\Delta_q$ with $q=4,5,\ldots$ can {\it not} be mapped onto the
percolation within our approach (see Sec.~\ref{s4.3} and Appendix). 

We now proceed by proving the statement 1. It is convenient for us to
recall first the corresponding proof for the case of a single Green
function, $\langle {\rm Tr} G(e,e;z)\rangle$, considered in
\cite{beamond02}. For an arbitrary edge $f$ the paths entering (\ref{e7a}) 
can be classified according to the number $k$ of times they pass through
$f$. The contribution of paths with $k\ne 0$ has the form
\be
\label{e21}
\sum_{k=1}^\infty \langle {\rm Tr} B[U_fA(f,f)]^k\rangle,
\ee
where $B_1$ is a sum over all paths going from $f$ to $e$ and then from $e$
to $f$, and $A(f,f)$ denotes a sum over paths which begin and end on
$f$ and do not return to $f$ in between. Since $A(f,f)$ 
is a linear combination of SU(2) matrices with real coefficients, it can be
represented as $A(f,f)=|A(f,f)|\tilde{A}(f,f)$, where $\tilde{A}(f,f)\in
SU(2)$ and $|A(f,f)|$ is a real number. After a change of the
integration variable, $U_f\tilde{A}(f,f)\to U_f$, Eq.~(\ref{e21}) then
reduces to
\be
\label{e22}
\sum_{k=1}^\infty\langle{\rm Tr} B U_f^k \rangle |A(f,f)|^k.
\ee
Since $SU(2)$ matrices can be represented as $U=\exp(i\alpha {\bf
n}\bbox{\sigma})$, with a real $\alpha$ and a unit vector ${\bf n}$
($\sigma_i$ are the Pauli matrices), one finds  
\be
\label{e23}
U^k=\cos k\alpha\cdot {\bf 1} + i\sin k\alpha\: {\bf n}\bbox{\sigma}.
\ee
The SU(2) invariant measure is $(2/\pi)\int_0^\pi d\alpha \sin^2\alpha \int
d{\bf n}$, where $d{\bf n}$ is the conventional measure on the sphere.
Therefore, for an integer $k$
\be
\label{e24}
\langle U^k\rangle =c_k\cdot {\bf 1}, \qquad 
c_k=\left\{\begin{array}{rl} 1,            & \ \ k=0 \\
                            -{1\over 2},  & \ \ k=2,-2\\
			    0,            & \ \ {\rm otherwise.}
\end{array}\right.
\ee
Substituting (\ref{e24}) in (\ref{e22}), one finds that only the term with
$k=2$ survives, which completes the proof of the the statement 1 for the
case of an average of one Green function. 

We turn now to the products of two Green functions. Consider
\be
\label{e25}
\langle {\rm Tr} G(e,e';z)G(e'',e''';z)\rangle
\ee
(we will need below both cases $e''=e'$, $e'''=e$, and $e''=e$, $e'''=e'$).
Using (\ref{e7a}), we classify the contributions to (\ref{e25}) according
to the numbers of returns $k_1$, $k_2$ to the edge $f$ for the
corresponding two paths. We want to show that only the contributions with
$k_1+k_2=0,\ 2$ are to be taken into account. If one of $k_i$ is zero, the
proof is obtained in the same way as for a single Green function (see
above). We thus consider the remaining contributions, which are of the
following form:
\be
\label{e26}
\sum_{k_1,k_2=1}^\infty\langle{\rm Tr} B_1[U_f A(f,f)]^{k_1} B_2 [U_f
A(f,f)]^{k_2}\rangle,
\ee
where $B_1$ is a sum over the paths $f\to e$ and $e'''\to f$, and $B_2$ is
a sum over the paths $f\to e''$ and $e'\to f$. Performing the variable
change $U_f\tilde{A}(f,f)\to U_f$ as before, we get
\be
\label{e27}
\sum_{k_1,k_2=1}^\infty\langle{\rm Tr} B_1 U_f^{k_1} B_2 U_f^{k_2}\rangle
|A(f,f)|^{k_1+k_2}.
\ee
Using (\ref{e23}), we calculate now the average over $U_f$ in (\ref{e27}):
\bea
\label{e28}
\langle {\rm Tr} B_1 U_f^{k_1} B_2 U_f^{k_2}\rangle & = & {\rm Tr} B_1 B_2
\langle\cos k_1\alpha\cos k_2\alpha\rangle_\alpha \nonumber\\
 &-& {1\over 3} {\rm Tr}
\sum_i B_1\sigma_i B_2\sigma_i \langle\sin k_1\alpha\sin
k_2\alpha\rangle_\alpha  \nonumber \\
& = & {1\over 2} {\rm Tr} (B_1 B_2 + {1\over 3} \sum_iB_1\sigma_i
B_2\sigma_i) c_{k_1+k_2} \nonumber \\
 &+& {1\over 2} {\rm Tr} (B_1 B_2 - {1\over 3} 
\sum_iB_1\sigma_i B_2\sigma_i) c_{k_1-k_2}  \nonumber \\
& \equiv & b_1 c_{k_1+k_2} + b_2 c_{k_1-k_2}.
\eea
The only property of the factors $b_1$, $b_2$ which is important for us at
this stage is that they are independent of $k_1$, $k_2$.
The sum (\ref{e27}) is therefore reduced to the form 
\be
\label{e29}
\sum_{k_1,k_2=1}^\infty(b_1c_{k_1+k_2}+b_2c_{k_1-k_2})|A(f,f)|^{k_1+k_2}. 
\ee
While the first term in brackets is non-zero only for $k_1+k_2=2$ ({\it
i.e.} $k_1=k_2=1$) as required, the second one seems to spoil the proof.
Let us perform, however, a summation over $k_1$ at fixed $k_1+k_2=k$. Using
Eq.~(\ref{e24}), we find then that the coefficients in the second term
cancel for any even $k\ge 4$ (for odd $k$ all terms are trivially zero):
\bea
\label{e30}
\sum_{k_1+k_2=k} c_{k_1-k_2} &=& c_{k-2}+c_{k-4}+\ldots +c_{-(k-2)} \nonumber \\
&=& c_2+c_0+c_{-2}=0.
\eea
Therefore, only the term with $k_1=k_2=1$ survives in the sum (\ref{e29}),
which completes the proof.\footnote{The correlation function
$\langle{\rm Tr}G(e,e;z){\rm Tr}G(e',e';z)\rangle$ is analyzed in the
same way, yielding again a sum of the type (\ref{e29}), so that our
argument remains valid.}
 
Applying now the statement 2, the proof of which is given in Appendix,
we represent each node as a superposition of contributions of the
types 
(i) and (ii) (Fig.~\ref{fig2}) with weights $\cos^2\theta$ and
$\sin^2\theta$, equal to 1/2 at the SQH critical point. The network
is then reduced to a weighted sum over all its possible decompositions
in a set of closed loops (such that each edge belongs to exactly one
loop). These loops can be viewed \cite{gruzberg99,beamond02} as hulls
of the bond percolation problem. Non-zero contributions to the
correlation function (\ref{e25}) come from pairs of paths retracing
exactly twice a loop or a part of it.  
This yields for $z<1$
\bea
\langle{\rm Tr} G(e',e;z)G(e,e';z)\rangle & = &
\langle{\rm Tr} G(e',e;z^{-1})G(e,e';z^{-1})\rangle \nonumber \\ 
&=& -2\sum _N P(e',e;N)z^{2N}, \label{e8} \\
\langle{\rm Tr} G^2(e',e;z)\rangle &=&
\langle{\rm Tr} G^2(e,e';z^{-1})\rangle \nonumber\\
&=& -\sum_N P_1(e',e;N)z^{2N}, \label{e9a} 
\eea
where $P(e',e;N)$ and $P_1(e',e;N)$ are probabilities that the edges
$e$ and $e'$ belong to the same loop of the length $N$ (resp. with the
length $N$ of the part corresponding to the motion from $e$ to
$e'$). Furthermore, to calculate the correlation function 
$\langle{\rm Tr} G(e',e;z)G(e,e';z^{-1})\rangle$ entering (\ref{e6}),
we apply the identity (\ref{e7b}) to the second Green function and then
use the property 
\be
\langle {\rm Tr} G(e',e;z)G^\dagger(e',e;z)\rangle = -2 
\langle {\rm Tr} G^2(e',e;z)\rangle
\label{e31}
\ee
following from the SU(2) symmetry. As a result, we find 
\be
\langle{\rm Tr} G(e',e;z) G(e,e';z^{-1})\rangle =
-2\sum_N P_1(e',e;N)z^{2N}, \label{e9} 
\ee
and, combining (\ref{e8}) and (\ref{e9}),
\bea
\label{e32}
\pi^2 D(e',e;\gamma) &=& {1\over 2} \sum_N [P_1(e',e;N) +
P_1(e,e';N)] z^{2N} \nonumber\\  
& - &\sum_N P(e',e;N)z^{2N}.
\eea

Equations (\ref{e8}), (\ref{e9}), (\ref{e32}) express the quantum
correlation functions entering (\ref{e6}) in terms of purely classical
quantities $P(e',e;N)$ and  $P_1(e',e;N)$. To analyze the results, we
recall some facts from the percolation theory. It is known that the
fractal dimension of the percolation 
hulls is 7/4 \cite{saleur87}, implying (see \cite{moore02} for a
recent discussion) that $P$ and $P_1$ scale as
\be
P(e',e,N),\ P_1(e',e,N) \sim N^{-8/7}r^{-1/4}\ , \qquad r\lesssim N^{4/7}
\label{e10}
\ee
and fall off exponentially fast at $r\gg N^{4/7}$, where $r$ is the distance
between $e$ and $e'$. This yields for the correlation functions in
(\ref{e8}) and (\ref{e9}) (which we abbreviate as 
$\langle G_R G_R\rangle$, $\langle G_A G_A\rangle$, 
$\langle G_R G_A\rangle$) 
\bea
&& \langle G_R G_R\rangle = \langle G_A G_A\rangle \simeq
\langle G_R G_A\rangle \sim r^{-1/2}, \nonumber\\
&& \hspace*{3cm} r\ll \xi_\gamma\equiv 
\gamma^{-4/7}
\label{e12}
\eea
in full agreement with the scaling argument of \cite{gruzberg99}. 
However, these leading order terms cancel in (\ref{e32}) since
\be
\label{e33}
\sum_N P(e',e,N) = \sum_N P_1(e',e,N)= P(e',e),
\ee
where $P(e',e)$ is the probability that the edges $e$ and $e'$ belong
to the same loop. The result is 
non-zero due to the factors $z^{2N}$ only, implying that relevant $N$
are now $N\sim \gamma^{-1}$, so that $\langle(G_R-G_A)(G_R-G_A)\rangle$
scales differently compared to (\ref{e12}),
\bea
{\cal D}(e',e;\gamma) & = & {1\over\pi^2}\sum_N
[P(r,N)-P_1(r,N)](1-e^{-2N\gamma}) \nonumber\\
&\sim & P(r,\gamma^{-1})\gamma^{-1}\sim
(\xi_\gamma r)^{-1/4},\qquad r\lesssim \xi_\gamma.
\label{e13}
\eea
Using now the definition (\ref{e4}) of ${\cal D}$ and the DOS scaling,
$\rho(\epsilon)\sim \epsilon^{1/7}\sim \xi_\epsilon^{-1/4}$, we find
for $r\lesssim \xi_\epsilon$
\be
\label{e14}
L^4\langle
\psi_{i\alpha}^*(e)\psi_{j\alpha}(e)\psi_{i\beta}(e')\psi_{j\beta}^*(e')
\rangle
\sim (r/\xi_\epsilon)^{-1/4}.
\ee

The correlation function (\ref{e7}) is calculated in a similar
way. The results for the $\langle G_RG_R\rangle$, $\langle G_AG_A\rangle$, 
and $\langle G_RG_A\rangle$ terms in (\ref{e7}) have the form 
\bea
&& \langle{\rm Tr} G(e,e;z) {\rm Tr} G(e',e';z)\rangle \nonumber\\ 
&& = 4-2\sum_N[P(e;N)+P(e';N)]z^{2N} \nonumber \\
&& + \sum_{NN'} P_-(e,e';N,N')z^{2(N+N')}+\sum_N P(e,e';N)z^{2N},
\label{e34a}\\
&& \langle{\rm Tr} G(e,e;z^{-1}) {\rm Tr} G(e',e';z^{-1})\rangle \nonumber \\
&& = \sum_{NN'} P_-(e,e';N,N')z^{2(N+N')}+\sum_N P(e,e';N)z^{2N},
\label{e34b}\\
&&\langle{\rm Tr} G(e,e;z) {\rm Tr} G(e',e';z^{-1})\rangle = 
2\sum_N P(e';N)z^{2N} \nonumber \\
&& 
- \sum_{NN'} P_-(e,e';N,N')z^{2(N+N')}-\sum_N P(e,e';N)z^{2N},
\label{e34c}
\eea
where $P(e;N)$ is the probability that $e$ belongs to a loop of the
length $N$, while $P_-(e,e';N,N')$ is the probability that $e$ and
$e'$ belong to different loops of the length $N$ and $N'$,
respectively. A larger number of terms in (\ref{e34a})--(\ref{e34c})
as compared to (\ref{e8}), (\ref{e9}) is because of two
reasons. First, there is a unit matrix contribution of a ``path of
zero length'' to the expansion (\ref{e7a}) of the Green function
$G(e,e;z)$. Second, $e$ and $e'$ may now belong to different loops and
still give a finite contribution, since each of the two paths will
retrace twice the corresponding loop. 

Combining (\ref{e34a})--(\ref{e34c}) and using the identities
\bea
&& P(e,e';N)+\sum_{N'}P_-(e,e;N,N')= P(e;N),
\label{e35a} \\
&& \sum_N P(e;N)=1,
\label{e35b}
\eea
we get for the correlation function (\ref{e7})
\bea
&& \pi^2\tilde{\cal D}(e',e;\gamma)=\sum_N P(e,e';N)(1-z^{2N}) \nonumber \\
&& + \sum_{NN'}  P_-(e,e';N,N')(1-z^{2N})(1-z^{2N'}). \label{e36}
\eea
We see again that at $z\equiv e^\gamma =1$ the result is zero, and
that at small $\gamma$ it is dominated by $N \sim\gamma^{-1}$. 
Using Eq.~(\ref{e10}) and 
\be
\label{e37}
P_-(e,e';N,N') \sim P(e;N) P(e';N') \sim N^{-8/7}(N')^{-8/7},
\ee
we find that the first term in (\ref{e36}) is $\sim
\xi_\gamma^{-1/4}r^{-1/4}$ at $r\ll \xi_\gamma$,  while the second one
is $\sim\xi_\gamma^{-1/2}$ and thus can be neglected. Therefore, we
find that $\tilde{\cal D}$ shows the same scaling behavior as ${\cal D}$
[see Eq.~(\ref{e13})],
\bea 
\tilde{\cal D}(e',e;\gamma) &\simeq& \sum_N
P(e,e';N)(1-e^{-2N\gamma})\nonumber\\ 
&\sim& \xi_\gamma^{-1/4}r^{-1/4},\qquad r\ll\xi_\gamma. 
\label{e38}
\eea
In other words, the wave function correlator $\langle|\psi_{i\alpha}^2(e)
\psi_{j\beta}^2(e')|\rangle$ with $\epsilon_i,
\epsilon_j\sim\epsilon$ scales at $r\lesssim \xi_\epsilon$ 
in the same way as (\ref{e14}),
\be
\label{e39}
L^4\langle|\psi_{i\alpha}^2(e)\psi_{j\beta}^2(e')|\rangle
\sim \rho^{-2}(\epsilon)
\tilde{\cal D}(e',e;\gamma\sim\epsilon) \sim(r/\xi_\epsilon)^{-1/4}.
\ee
Both Eqs. (\ref{e14}) and (\ref{e39}) imply that the fractal exponent
\be
\label{e39a}
\eta\equiv -\Delta_2= {1\over 4},
\ee
at variance with what one might naively expect
from the $r^{-1/2}$ scaling of the diffusion propagator 
$\langle G_RG_A\rangle$,  Eq.~(\ref{e12}).

\subsection{Three-point functions}
\label{s4.2}

We consider now averaged products of three Green functions, analogous
to the two-point functions (\ref{e6}) and (\ref{e7}),
\bea
\label{e40}
{\cal D}(e,e',e'';\gamma)&=&(2\pi)^{-3}
\langle{\rm Tr}[G(e,e';z)-G(e,e';z^{-1})] \nonumber \\
&\times & [G(e',e'';z)-G(e',e'';z^{-1})] \nonumber \\
&\times & [G(e'',e;z)-G(e'',e;z^{-1})]
\rangle, \\
\tilde{\cal D}(e,e',e'';\gamma)&=&(2\pi)^{-3}
\langle{\rm Tr}[G(e,e;z)-G(e,e;z^{-1})] \nonumber \\
&\times &{\rm Tr} [G(e',e';z)-G(e',e';z^{-1})] \nonumber \\
&\times &{\rm Tr} [G(e'',e'';z)-G(e'',e'';z^{-1})]
\rangle.
\label{e41}
\eea
The key role in the calculation of (\ref{e40}) and (\ref{e41}) is
played by the proofs of applicability of the statements 1 and 2
(Sec.~\ref{s4.1}) to the products of three Green functions. Details of
these proofs are given in Appendix. After the two statements are
applied and the network is reduced to a sum over its loop
decompositions (as in Sec.~\ref{s4.1}), the correlation functions are
calculated straightforwardly. In particular, we find for the averaged
products of three Green functions entering (\ref{e40})
\bea
&& \langle{\rm Tr} G(e,e';z)G(e',e'';z)G(e'',e;z)\rangle \nonumber \\
&& = -\sum_N[3P(e,e',e'';N)+P(e'',e',e;N)]z^{-2N},
\label{e42} \\
&& \langle{\rm Tr}
G(e,e';z^{-1})G(e',e'';z^{-1})G(e'',e;z^{-1})\rangle 
\nonumber \\
&& = \sum_N[P(e,e',e'';N)+3P(e'',e',e;N)]z^{-2N},
\label{e43} \\
&& \langle{\rm Tr} G(e,e';z)G(e',e'';z)G(e'',e;z^{-1})\rangle \nonumber \\
&& = -2\sum_N P_1(e,e',e'';N)z^{-2N},
\label{e44} \\ 
&& \langle{\rm Tr} G(e,e';z^{-1})G(e',e'';z^{-1})G(e'',e;z)\rangle \nonumber \\
&& = 2\sum_N P_1(e'',e',e;N)z^{-2N},
\label{e45}
\eea
where $P(e,e',e'';N)$ is the probability that the edges $e$, $e'$ and
$e''$ belong to the same loop of the length $N$, with $e'$ lying on
the path from $e''$ to $e$, while  $P_1(e,e',e'';N)$ is the same
probability but with $N$ being the length of the segment from $e''$ to
$e$. Combining Eqs.~(\ref{e42})--(\ref{e45}), we express the
correlation function (\ref{e40}) in terms of the classical
probabilities $P$ and $P_1$. Remarkably, the situation is
qualitatively different as compared to the calculation of two-point
functions (Sec.~\ref{s4.1}): the leading terms in
(\ref{e42})--(\ref{e45}) do {\it not} cancel in the expression for
${\cal D}(e,e',e'';\gamma)$. We can thus simply set $\gamma=0$ ($z=1$),
which yields
\bea
\label{e46}
(2\pi)^3{\cal D}(e,e',e'';\gamma)\simeq
2[P(e,e',e'')+P(e'',e',e)],&& \nonumber \\
r\ll\xi_\gamma, &&
\eea
where $P(e,e',e'')=\sum_N P(e,e',e'';N)$ is the probability for
$e$, $e'$, and $e''$ to belong to the same loop with the orientation
$e\leftarrow e'\leftarrow e''\leftarrow e$, and $r$ is the
characteristic scale of the distances between $e$, $e'$, and
$e''$.

The correlation function (\ref{e41}) is calculated in the same way,
and the results are qualitatively similar. We thus skip 
intermediate formulas and only present the final result,
\bea
\label{e47}
(2\pi)^3\tilde{\cal D}(e,e',e'';\gamma)\simeq
8[P(e,e',e'')+P(e'',e',e)],&& \nonumber \\
\qquad r\ll\xi_\gamma,&&
\eea
which differs from (\ref{e46}) by an overall factor of 4 only. 

Using any of the equations (\ref{e46}), (\ref{e47}), we can determine
the fractal exponent $\Delta_3$. In analogy with (\ref{e10}), the
probability for the edges $e$, $e'$, and $e''$ separated by distances
$\sim r$ to belong to the same loop (percolation hull) of a length $N$
scales as 
\be
\label{e48}
P(e,e',e'';N)\sim N^{-8/7}r^{-1/2}, \qquad
r\lesssim N^{4/7} 
\ee
and is exponentially small for $r\gg N^{4/7}$. Summing over $N$, we
thus get
\be
P(e,e',e'')\sim r^{-3/4}. 
\label{e49}
\ee
Substituting this in Eqs.~(\ref{e46}), (\ref{e47}) and expressing
${\cal D}$ and $\tilde{\cal D}$ in terms of wave functions in analogy
with the two-point functions (\ref{e4}), (\ref{e6a}), we find for
$r\lesssim \xi_\epsilon$
\bea
\label{e50}
&& L^6\langle
\psi_{i\alpha}(e)\psi_{i\beta}^*(e')\psi_{j\beta}(e')
\psi_{j\gamma}^*(e'')\psi_{k\gamma}(e'')\psi_{k\alpha}^*(e)
\rangle, \nonumber \\
&& L^6\langle|\psi_{i\alpha}(e)\psi_{j\beta}(e')\psi_{k\gamma}(e'')|^2\rangle
\sim{r^{-3/4}\over\rho^3(\epsilon)} \sim (r/\xi_\epsilon)^{-3/4}.
\eea
Therefore, the exponent $\Delta_3$ is equal to 
\be
\label{e51}
\Delta_3=-{3\over 4}.
\ee

\subsection{Discussion}
\label{s4.3}

The situation we encountered while calculating two- and three-point
functions is qualitatively different from what happens at conventional
localization transitions. Specifically, in the conventional case
average products of only retarded or only advanced Green functions are
negligible compared to mixed averages containing both $G_R$ and $G_A$,
{\it e.g.} $\langle G_RG_R\rangle,\ \langle G_AG_A\rangle \ll
\langle G_RG_A\rangle$. For this reason, the wave function
correlators, which are proportional to 
$\langle(G_R-G_A)(G_R-G_A)\rangle$, are determined by $\langle
G_RG_A\rangle$ (and similarly for higher moments). In contrast, we
have found in the SQH case that the correlators of the 
$\langle G_RG_R\rangle$ (or $\langle G_AG_A\rangle$) type are
approximately equal to $\langle G_RG_A\rangle$ and cancel it in the
leading order (so that $\langle(G_R-G_A)(G_R-G_A)\rangle$ scales
differently). Evaluation of three-point functions made the overall
picture even more complex: while we obtained again an identical
scaling of, say, $\langle G_RG_RG_R\rangle$ and $\langle
G_RG_RG_A\rangle$  correlators, this time the cancelation was not
complete, and the correlation function
$\langle(G_R-G_A)(G_R-G_A)(G_R-G_A)\rangle$ scaled in the same way.

To shed more light on the reason for these different types of scaling
behavior, it is instructive to reverse the logic and to examine how
the diffuson scaling (\ref{e12}) can be
obtained from wave function correlations (\ref{e14}). It is
straightforward to express the zero-energy diffusion propagator 
in terms of the
correlation function ${\cal D}(e',e;\epsilon_1,\epsilon_2)$ defined in
Eq.~(\ref{e5}),
\bea
\label{e52}
&& \Pi(e',e)  \equiv  \langle {\rm Tr} G_R(e',e;1)G_A(e,e';1)\rangle
\nonumber \\
&&  =  \int{d\epsilon_1 d\epsilon_2 \over(1-e^{-i\epsilon_1+0})
(1-e^{-i\epsilon_2-0})}{\cal D}(e',e;\epsilon_1,\epsilon_2).
\eea  
As discussed in Sec.~\ref{s4.1}, ${\cal
D}(e',e;\epsilon_1,\epsilon_2)$ scales with the distance $r=|e'-e|$
and the energy $\epsilon_{1,2} \sim \epsilon$ as
follows
\be
\label{e53}
{\cal D}(e',e;\epsilon_1,\epsilon_2) \sim 
(r/\xi_\epsilon)^{\Delta_2}\xi_\epsilon^{-2x_\rho},\qquad
r\ll\xi_\epsilon,
\ee
where $\Delta_2=-1/4$, $x_\rho=1/4$ is the scaling dimension of DOS
defined by $\rho(\epsilon)\sim\xi_\epsilon^{-x_\rho}$, and 
$\xi_\epsilon=\epsilon^{-1/(2-x_\rho)}=\epsilon^{-4/7}$. (For
$r\gg\xi_\epsilon$ ${\cal D}(e',e;\epsilon_1,\epsilon_2)$ is
exponentially small.)

Substituting (\ref{e53}) in (\ref{e52}), we see that if
$2x_\rho+\Delta_2>0$, which is the case for the SQH transition, the
energy integral in (\ref{e52}) is dominated by
$\epsilon_{1,2}\sim\epsilon(r)$, where $\epsilon(r)$ is defined by
$\xi_{\epsilon(r)}\sim r$ ({\it i.e.} $\epsilon(r)\sim
r^{-(2-x_\rho)}=r^{-7/4}$), and can be estimated as
\bea
\label{e54}
\Pi(e',e) & \sim & {\cal
D}(e',e;\epsilon_1,\epsilon_2)|_{\epsilon_{1,2}\sim \epsilon(r)}
\nonumber\\
& \sim & r^{-2x_\rho} =r^{-1/2},
\eea  
in full agreement with the exact result (\ref{e12}).\footnote{Since the
integral (\ref{e52}) is determined by the upper cutoff $\epsilon(r)$
(and not by the vicinity of $\epsilon=0$), this calculation applies
not only to $\Pi=\langle G_RG_A\rangle$, but equally well to $\langle
G_RG_R\rangle$  and $\langle G_AG_A\rangle$, in agreement with
(\ref{e12}).} This is in a stark contrast with the case of a
conventional localization (Anderson or QH) transition, when the
diffusion propagator $\Pi$ (or any other correlation function of the
$\langle G_RG_A\rangle$ type) depends in a singular way on the
infrared cutoff set by $L_\omega$,
see the last line of 
Eq.~(\ref{e2}). On the other hand, Eq.~(\ref{e54}) has a familiar
form of a two-point correlator in a conformal field theory (or, more
generally, in field-theoretical description of standard critical
phenomena), where
$\langle {\cal O}_i(\br_1){\cal O}_i(\br_2)\rangle$ scales as
$|\br_1-\br_2|^{-2x_i}$, with $x_i$ being the scaling dimension of the
operator ${\cal O}_i$. 

Generalization to higher moments  is straightforward. We define a wave
function correlation function
\bea
\label{e55}
&& {\cal D}^{(q)}(e_1,\ldots,e_q;\epsilon_1,\ldots,\epsilon_q)
=(2\pi)^{-q} \nonumber \\
&&\times \langle{\rm Tr}[(G_R-G_A)(e_1,e_2;e^{i\epsilon_1}) 
 (G_R-G_A)(e_2,e_3;e^{i\epsilon_2}) \nonumber\\ 
&& \times \ldots \times (G_R-G_A)(e_q,e_1;e^{i\epsilon_q})]\rangle,
\eea
and a set of $\langle G\ldots G\rangle$ correlation functions,
\bea
\label{e56}
&& \Pi^{(q)}_{s_1\ldots
s_q}(e_1,\ldots,e_q;\epsilon_1,\ldots,\epsilon_q)
\nonumber \\
&& =\langle{\rm Tr}G_{s_1}(e_1,e_2;e^{i\epsilon_1})
\ldots G_{s_q}(e_q,e_1;e^{i\epsilon_q})\rangle,
\eea
where $s_j=R$ or $A$. Assuming that all distances between the points
$e_i$ are $\sim r$, we have in analogy with (\ref{e53}),
\be
\label{e57}
{\cal D}^{(q)}(e_1,\ldots,e_q;\epsilon_1,\ldots,\epsilon_q)\sim
(r/\xi_\epsilon)^{\Delta_q} \xi_\epsilon^{-qx_\rho}, \qquad r\lesssim
\xi_\epsilon.  
\ee
Writing for $\Pi^{(q)}_{s_1\ldots s_q}$ a spectral representation
of the type (\ref{e52}), we see that the integrals are determined by
the upper limit $\epsilon\sim \epsilon(r)=r^{-(2-x_\rho)}$ provided
\be
\label{e58}
\Gamma(q) \equiv qx_\rho+\Delta_q> 0.
\ee
Under this condition, we find that  $\Pi^{(q)}_{s_1\ldots s_q}$ is in
fact independent of the indices $s_i$ and scales as 
\be
\label{e59}
\Pi^{(q)}_{s_1\ldots
s_q}(e_1,\ldots,e_q;\epsilon_1,\ldots,\epsilon_q)\sim r^{-qx_\rho}.
\ee
For larger $q$, when $qx_\rho+\Delta_q<0$, the energy integrals are
dominated by the vicinity of $\epsilon=0$. Consequently, the
correlation functions $\Pi^{(q)}_{s_1,\ldots,s_q}$ start to depend in
a singular way on the infrared cutoff ($\xi_\epsilon$) and are
expected to scale in the same way as ${\cal D}^{(q)}$,
Eq.~(\ref{e57}) (with a numerical prefactor depending on indices
$s_i$), similarly to the conventional Anderson localization
transition.

The value of $q$ separating the two regimes is thus determined by the
equation $qx_\rho+\Delta_q=0$. For the SQH transition ($x_\rho=1/4$)
its solution is, in view of Eq.~(\ref{e51}), $q=3$. Remarkably, this
is also the largest value of $q$ for which the mapping onto
percolation described above still works (see Appendix). 
We believe that this is not a
mere coincidence. Indeed, within this mapping  average products
$\Pi^{(q)}_{s_1\ldots s_q}$ of Green functions are expressed in terms
of probabilities of the percolation theory, 
and are therefore of order unity for $r \sim 1$. 
On the other hand, Eq.~(\ref{e57}) yields, in the regime
$qx_\rho+\Delta_q<0$, a result which is much larger than unity at
$r\sim 1$, $\xi_\epsilon\gg 1$ and diverges in the absence of the
infrared cutoff, $\xi_\epsilon\to\infty$. We see no way how such a
behavior might be produced by the percolation theory. 

Finally, we discuss a relation between our consideration and the
field-theoretical approach to the wave-function multifractality
\cite{wegner80,wegner85,duplantier91,mudry96,bhaseen00,bernard01b}.
In the renormalization-group language, $\Gamma(q)$ defined by
Eq.~(\ref{e58}) are scaling dimensions of operators of the type 
${\cal O}^{(q)}\sim \psi_{s_1}\psi^\dagger_{s'_1}\ldots
\psi_{s_q}\psi^\dagger_{s'_q} $, where $\psi,\psi^\dagger$ are
electronic fields. Averaged products of Green functions are expressed
as correlation functions of the corresponding operators ${\cal
O}^{(q)}$; in particular, (\ref{e56}) takes the form
\be
\label{e60}
\Pi^{(q)}_{s_1\ldots s_q} \sim \langle {\rm Tr} 
{\cal O}^{(1)}_{s_1s_2}(e_2) {\cal O}^{(1)}_{s_2s_3}(e_3) \ldots     
{\cal O}^{(1)}_{s_qs_1}(e_1)  \rangle.
\ee
To calculate the scaling behavior of such correlation functions, one
applies the operator product expansion (OPE)
\cite{wegner85,duplantier91,mudry96}. Generically, the identity
operator will be among those generated by  the OPE. Moreover, under
the condition $\Gamma(q)>0$ [Eq.~\ref{e58})] it will be the most
relevant operator and will dominate the expansion, leading to the gap
scaling $\Pi^{(q)}\sim r^{-q\Gamma(1)}$, in agreement with
(\ref{e59}). On the other hand, if $\Gamma(q)<0$, the operator 
${\cal O}^{(q)}$ will give a dominant contribution to OPE, leading to a
multifractal type of scaling, $\Pi^{(q)}\propto 
r^{-q\Gamma(1)}(r/\xi_\epsilon)^{\Gamma(q)}$, as in
Eq.~(\ref{e57}). What is, however, non-trivial from this point of
view, is that the scaling of the wave function correlator (\ref{e55})
has the multifractal form (\ref{e57}) independently of the sign of
$\Gamma(q)$. This means that in the regime $\Gamma(q)>0$ the leading
(gap scaling) terms (\ref{e59}) cancel in the particular combination
of the functions $\Pi^{(q)}$ corresponding to ${\cal D}^{(q)}$, and
subleading terms determine the result (\ref{e57}). A similar
cancellation of leading scaling terms in the context of classical
percolation was recently discussed in \cite{moore02}.

\section{Wave function statistics: numerical results}
\label{s5}

\subsection{Multifractality spectrum}
\label{s5.1}

The analytical treatment of Sec.~\ref{s4} yielded results for the
anomalous dimensions at two distinct values of $q$,
$\Delta_2=-1/4$ and $\Delta_3=-3/4$. In order to obtain more complete
information 
about the wave function statistics, namely the multifractality
spectrum at arbitrary $q$, we have performed numerical simulations.
A question we are particularly interested in is whether or not the
spectrum is exactly parabolic. A definite answer on this question
will imply, along with exact values of $\Delta_2$ and $\Delta_3$,  
an important constraint on the conformal theory of the
critical point, which is a subject of current research
\cite{fendley00,bernard01a,bernard01b}. 

Before we come to the presentation of our findings, we give a few remarks
about technical aspects of our numerics. We compute
wave functions by numerically diagonalizing the $4L^2 \times 4L^2$
unitary time evolution operator $\cal U$  of the
Chalker-Coddington network described in Sec.~\ref{s3}.
Using advanced sparse matrix packages \cite{numerics}, we 
selectively calculate only states with energies in the vicinity of
$\epsilon=0$, which are critical over the whole extent of the
system ($\xi_\epsilon\sim L$). 
Specifically, we consider, for each realization of the
network, four lowest eigenstates ({\it i.e.} with eigenvalues
$e^{-i\epsilon}$ closest to unity). The number of wave functions in a
statistical  
ensemble we obtain this way ranges from about $10^7$ for $L=16$ to 
$2\cdot10^4$ for $L=384$. 

\begin{figure}
\centerline{
\includegraphics[width=0.95\columnwidth]{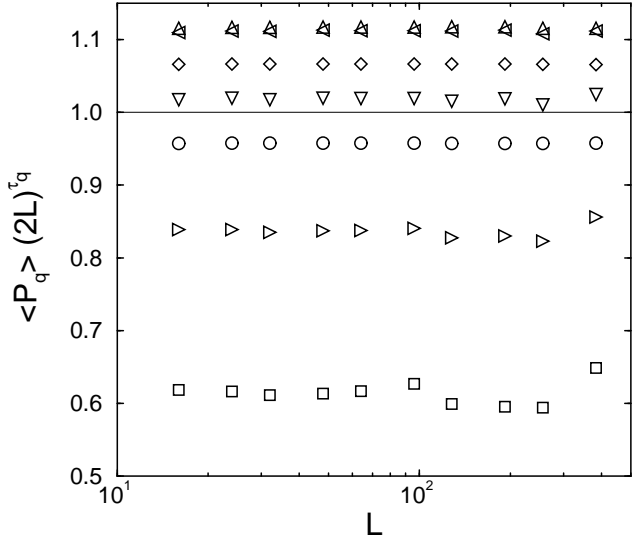}}
\caption{Scaling of the average IPR with the
system size $L$ for several values of $q=0.5 (\circ)$,  
$1.5 (\Diamond)$, $2 (\vartriangle)$, $2.5 (\triangleleft)$, 
$3(\triangledown)$, $3.5 (\triangleright)$, $4 ({\mbox{\tiny$\square $}})$. 
The system size dependence of the amplitude
$c_q(L)\equiv \langle P_q \rangle (2L)^{\tau_q}$ is presented, with
$\tau_q\equiv 2(q-1)+\Delta_q$ shown in Fig.~\ref{fig3}. 
The scattering of the data is due to the limited size 
of the statistical ensemble used.
The solid line is a guide to
the eye corresponding to the vanishing of finite size corrections
($c_q(L)={\rm const}$).}
\label{fig2a}
\end{figure}

To determine the multifractality spectrum $\tau_q$, we calculate for
each wave function $\psi_i$ the generalized inverse participation ratio (IPR) 
\be
\label{num1}
P_q = \sum_{\alpha e}|\psi_{i\alpha}(e)|^{2q}
\ee
and analyze the scaling of the average $\langle P_q\rangle$ with the
system size $L$. The data can be fitted very well by the power law
\be
\label{num2}
  \langle P_q \rangle = c_q (2L)^{-\tau_q}. 
 \ee
To demonstrate this, we show in Fig.~\ref{fig2a} the system size
dependence of $\langle P_q \rangle (2L)^{\tau_q}$, with $\tau_q$
obtained from the fit. 
The plot is organized in such a way that a pure power law (\ref{num2})
would correspond to a horizontal line. This kind of plot
is very sensitive to any corrections to a pure
power-law behavior of $\langle P_q\rangle$. Since no systematic curvature is
observed, corrections to scaling are extremely small.
This allows us to determine the anomalous dimensions 
$\Delta_q=\tau_q+2(1-q)$ with great accuracy.

\begin{figure}
\centerline{
\includegraphics[width=0.95\columnwidth]{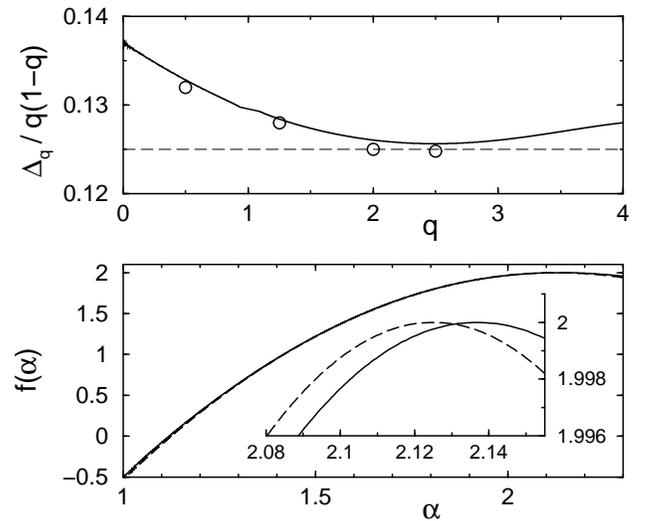}}
\vspace{3mm}
\caption{Upper panel:
Anomalous dimension $\Delta_{q}$ (solid line)
describing the scaling of the average IPR $\langle P_q\rangle$.
The functional form $\Delta_{q}/q(1-q)$
highlights the deviation from exact parabolicity (\ref{num3})
indicated by the dashed line. The circles correspond to the exponent
$\tilde{\Delta}_q$ obtained from the scaling of the typical value
$P_q^{\rm typ}$. \\
Lower panel: Singularity spectrum $f(\alpha)$. 
numerical results (solid line) and the parabolic approximation
(\ref{num4}) (dashed line) are shown. The inset depicts a
magnification of the apex region; the deviations from (\ref{num4})
correspond to the enhancement of $\Delta_q/q(1-q)$ 
near $q=0$ in the upper graph.}
\label{fig3}
\end{figure}

The obtained results for $\Delta_q$ are shown by a solid line in the
upper panel of Fig.~\ref{fig3}. We choose to plot $\Delta_q/q(1-q)$,
since this would give a constant for an exactly parabolic spectrum,
which is uniquely determined by $\eta$, $\Delta_q=\eta
q(1-q)/2$. According to our analytical calculations (Sec.~\ref{s4}), 
$\Delta_q/q(1-q)$ is equal to 1/8 for both $q=2$ and $q=3$; this value
is marked by the dashed line in the figure. It is seen that the
numerical results agree perfectly well with the analytical findings at
$q=2$ and $q=3$. Furthermore, the parabolic dependence may serve as a
numerically good approximation in the whole range of $q$ we studied,
\be
\label{num3}
\Delta_q\simeq {q(1-q)\over 8}.
\ee
Nevertheless, we believe that Eq.~(\ref{num3}) is {\rm not} exact.
Indeed, at $0 < q <2$ the numerically found $\Delta_q$ show clear
deviations from exact parabolicity (\ref{num3}), 
which are of the order of $10\%$
near $q=0$. Since this is precisely the regime in which finite-size 
effects have been found  to be very weak and $\Delta_q$ was determined
with a high accuracy, we interpret the 
observed deviations as very strong evidence for nonparabolicity
of the exact multifractal spectrum of the SQH transition.
In particular, the deviation of the limiting value 
$\Delta_q/q(1-q)|_{q\to 0}=0.137\pm 0.003$ from 1/8 well exceeds the
estimated numerical uncertainty.

We also calculated typical inverse participation ratios,
$P_q^{\rm typ} = \exp \langle \ln P_q \rangle$ and the
corresponding dimensions $\tilde{\tau}_q \equiv
2(q-1)+\tilde{\Delta}_q$.\footnote{In our earlier publications
\cite{adm-00,evers01,mildenberger02} we used the symbol $\tau_q$ to
characterize the scaling of the typical value $P_q^{\rm typ}$, and
$\tilde{\tau}_q$ for the average $\langle P_q\rangle$. In the present
paper we have chosen to interchange the notations.} 
It follows from the general analysis of the wave function
multifractality \cite{adm-00} that $\tilde{\tau}_q=\tau_q$ for 
$q\le q_c$, where $q_c$ corresponds to the zero $\alpha_-$ of the
singularity spectrum $f(\alpha)$ (defined below). In the present case
we find from the $\Delta_q$ data 
$q_c=3.9\pm 0.1$ (the parabolic approximation (\ref{num3})
would imply $q_c=4$). For $q>q_c$ the average $\langle P_q\rangle$ is
determined by rare realizations, and
$\tau_q<\tilde{\tau}_q$. Furthermore, already for $q$ smaller than but
close to $q_c$, finite-size corrections to $P_q^{\rm typ}$ become
large \cite{evers01}, leading to large errors in determination of
$\tilde{\tau}_q$. For the SQH problem, we find that the scaling of
$P_q^{\rm typ}$ exhibit small finite-size corrections as long $q\le
2.5$, so that the corresponding exponents $\tilde{\tau}_q$ 
can be found with a high accuracy. The results are shown by circles in
Fig.~\ref{fig3} (upper panel) and are in full agreement with the
values of $\tau_q$ obtained from the scaling of $\langle
P_q\rangle$. For larger $q$ ($q\ge 3$) the finite-size corrections to
$P_q^{\rm typ}$ (which we estimate  to be $\sim L^{-y}$ with $y
\approx 0.4$) become appreciable, strongly reducing the accuracy of
determination of $\tilde{\tau}_q$.

The lower panel of Fig.~\ref{fig3} depicts the singularity spectrum
$f(\alpha)$ obtained by a numerical Legendre transform
of the scaling dimension $\tau_q$, $f(\alpha_q)=q\alpha_q-\tau_q$ with
$\alpha_q=d\tau_q/dq$. The dashed line represents the parabolic approximation
corresponding to (\ref{num3}),
\be
\label{num4}
f(\alpha)=2-{(\alpha-\alpha_0)^2\over 4(\alpha_0-2)},\qquad
\alpha_0-2=1/8.
\ee
We see again that the parabolic approximation is numerically rather
good; nevertheless, it is not exact. Deviations from (\ref{num4}) are
demonstrated in the inset which shows an enlarged view of a region
around the maximum $\alpha_0$ of $f(\alpha)$. The deviation of
$\alpha_0-2=0.137\pm 0.003$ from 1/8 corresponds  to non-parabolicity
of $\tau_q$ discussed above.

\subsection{IPR fluctuations}
\label{s5.2}

\begin{figure}
\centerline{
\includegraphics[width=0.85\columnwidth]{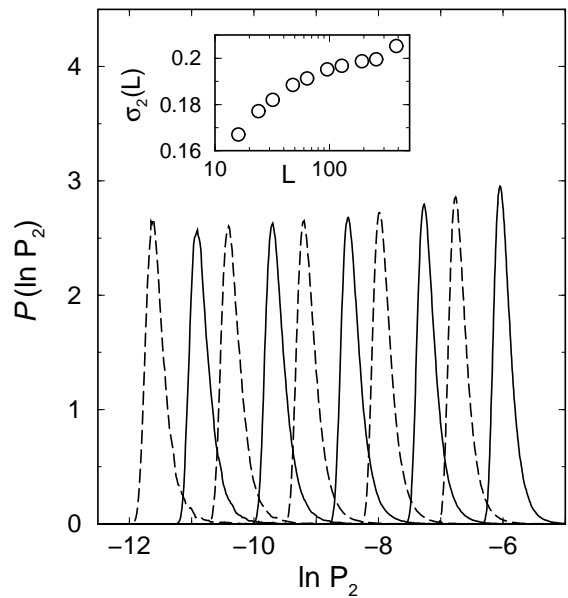}}
\vspace{3mm}
\caption{Distribution function ${\cal P}(\ln P_2)$ for system sizes
$L=16,\, 24,\, 32,\, 48,\, 64,\, 96,\, 128,\, 192,\, 256,\, 384$ 
(from right to left).
For values of $L$ large enough, the form
of  the distribution becomes independent of $L$.
Inset: Width $\sigma_2(L) = \langle ( \ln P_2 - \langle \ln P_2 \rangle )^2
\rangle^{1/2}$ of the distribution ${\cal P}(\ln P_2)$ versus
the system size $L$. Some scattering of the data for the largest
system sizes is due to the limited number of samples.}
\label{fig5}
\end{figure}

\begin{figure}
\centerline{
\includegraphics[width=0.95\columnwidth]{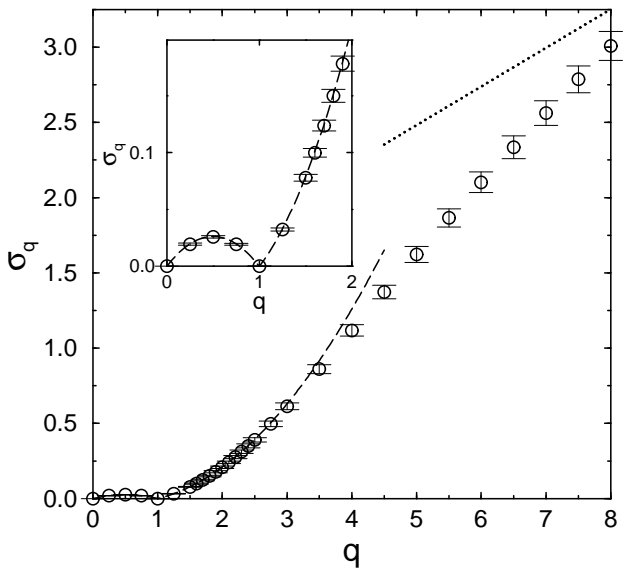}}
\vspace{3mm}
\caption{R.m.s. deviation $\sigma_q=\langle (\ln P_q-\langle \ln P_q\rangle)^2
\rangle^{1/2}$ of the IPR logarithm. 
The dashed line is a fit to Eq.~(\ref{ipr-fluc}),
the dotted line corresponds to the asymptotic limit
$\sigma_q=q/q_c$ with $q_c=3.9$.
The inset shows an enlarged view of the low-$q$ region.}
\label{fig6}
\end{figure}

We devote the remainder of this section to a brief discussion of 
the IPR distribution function ${\cal P}(P_q)$, specifically, its
evolution with the system size $L$ and dependence on $q$. In analogy
with Anderson and quantum Hall transitions studied earlier
\cite{adm-00,evers01,cuevas02a,mildenberger02,cuevas02b},  we expect
the distribution ${\cal P}(P_q)$ to become scale-invariant in the
large-$L$ limit. Figure \ref{fig5} demonstrates that this is indeed
the case. It represents the evolution of the distribution of $\ln P_2$
with the system size $L$. The mean of the distribution is shifted as
$-\tilde{\tau}_2\ln L$. Apart from small statistical fluctuations at
the largest system sizes, a clear tendency towards an asymptotic form
is observed. To characterize the width of the distribution
${\cal P}(\ln P_q)$, we calculate the variance $\sigma_q^2={\rm
var}(\ln P_q)$, as shown in the inset of Fig.~\ref{fig5} for
$q=2$. The results extrapolated to $L\to\infty$ (the finite-size
corrections are again of the type $L^{-y}$ with $y\approx 0.4$) are
presented in Fig.~\ref{fig6}. The behavior of $\sigma_q$ is
qualitatively similar to that found for other localization
transitions. A somewhat unusual feature of the SQH transition is that
in a rather broad range $0\le q\le 3$ the variance $\sigma_q^2$ is
remarkably well 
described by the formula 
\be
\label{ipr-fluc}
\sigma^2_q={\rm const}\times q^2(q-1)^2,
\ee
which has been derived for a metallic system
\cite{fyodorov95,adm-review}, or for
the Anderson transition with a weak multifractality, {\it e.g.} 
in $2+\epsilon$ dimensions \cite{adm-00,mildenberger02}. 
In the latter case this formula is valid for $q\ll q_c$. 
The accuracy of Eq.~(\ref{ipr-fluc})  is one more manifestation of the 
``close-to-metal'' character of the SQH critical point already
mentioned in Sec.~\ref{s3}, which leads to a relatively large value of
$q_c\simeq 4$. At larger $q$, the behavior of $\sigma_q$ 
becomes linear (as was also found for the conventional Anderson
transition \cite{mildenberger02,cuevas02b}), in agreement with the
theoretical prediction 
$\sigma_q=q/q_c$ for $q\gg q_c$ \cite{mildenberger02}. 
This is because in this regime the distribution ${\cal P}(P_q)$ is
dominated by a slowly decaying power-law tail ${\cal P}(P_q)\propto
P_q^{-1-x_q}$, where $x_q=q_c/q$ for $q>q_c$ \cite{adm-00}.

\section{Statistics of two-point conductances}
\label{s6}

So far, we have investigated properties of an open system. To define
the two-terminal conductance $g$, one opens the system by attaching
two leads. According to the Landauer-B\"uttiker formula, 
$g={\rm Tr} t^\dagger t$, where $t$ is the transmission matrix between
the leads. In the framework of network models, the transmission matrix
determining the two-point conductance between the edges $e$ and $e'$
has the form \cite{janssen99}
\be
\label{cond1}
t=\langle e'|(1-{\cal U}P)^{-1}{\cal U}|e\rangle,
\ee
where $P={\bf 1}-|e\rangle\langle e|-|e'\rangle\langle e'|$ projects
out the states on the edges $e$ and $e'$. 

Statistics of the two-point conductance $g(\br,\br')$ has been
extensively studied, both analytically and numerically, for the
conventional QH transition, exemplifying a localization transition
with non-critical DOS. It was shown that the moments 
$\langle g^q(\br',\br)\rangle$ obey a power-law scaling
\cite{zirnbauer94,janssen99,zirnbauer99}, 
\be
\label{cond1a}
\langle g^q(\br,\br')\rangle \sim |\br-\br'|^{-X_q},
\ee 
with a set of exponents $X_q$ related to $\Delta_q$
\cite{klesse01,evers01},    
\be
\label{e3}
X_q=\left\{ \begin{array}{ll}
\Delta_q+\Delta_{1-q}\ , & \qquad q<1/2 \\
2\Delta_{1/2}        \ , & \qquad q>1/2. 
\end{array} \right. 
\ee       
For the SQH  critical point, only the average conductance $\langle
g(e',e)\rangle$ has been considered previously. Gruzberg, Ludwig, and
Read \cite{gruzberg99} found that
\be
\label{cond2}
X_1=2 x_\rho = {1\over 2}.
\ee
Beamond, Cardy, and Chalker \cite{beamond02} used the mapping onto
percolation to calculate $\langle g(e',e)\rangle$ at the
band center $\epsilon=0$ ($z=1$), with the result
\be
\label{cond3}
\langle g(e',e)\rangle  = 2 P(e',e),
\ee
with $P(e',e)$ as defined in Eq.~(\ref{e33}). Comparing (\ref{cond3})
with (\ref{e9}), we see that $\langle g(e',e)\rangle$ is equal (up to
the sign) to the diffusion propagator $\Pi(e',e)=\langle {\rm Tr}
G_R(e',e;1)G_A(e,e';1)\rangle$. 

In this section, we will study statistical properties of $g(e',e)$ at
the SQH transition. Note that though the definition of $\langle
g(e',e)\rangle$ reminds closely that of the diffusion
propagator $\Pi(e',e)$, the identical scaling of the both quantities
is not at all self-evident. In contrast, they scale differently at
conventional localization transitions, as can be easily seen by
comparing Eqs.~(\ref{cond1a}), (\ref{e3}) with (\ref{e2}). It is
worthwhile to remind the reader the physical reason for this
difference (see also a related discussion in \cite{janssen99}). The product 
$\langle G_E^R(\br',\br)G_E^A(\br,\br')\rangle$ has a meaning of the
particle density (or, in an optical analogy, the radiation intensity) 
at a point $\br'$ induced by a source inserted
into the system at a point $\br$. In an infinite system at criticality
this quantity turns out to be infrared divergent: if a source is
switched on at a time $t=0$, the detected intensity will increase with
time without saturation, since the radiation cannot propagate away fast
enough. Therefore, in order to make $\langle G^RG^A\rangle$ finite,
one needs to allow the propagating wave to get out of the system,
{\it i.e.} to introduce absorption. One possibility is to make the
absorption weak but uniform over the whole system , leading to 
$\langle
G^R_{E+i\gamma}(\br',\br)G^A_{E-i\gamma}(\br,\br')\rangle\equiv
\Pi(\br',\br;2i\gamma)$, which is the same as introducing a small
uniform level broadening $\gamma$ (or equivalently, a small
frequency $\omega$ with an analytical continuation to the imaginary
axis, $\omega=2i\gamma$). Alternatively, one can allow for a
particle to be absorbed at the points $\br$ and $\br'$ only, but with
a probability of order unity, yielding the two-point conductance 
$g(\br',\br)$. Clearly, two definitions are essentially different
(which is already obvious from the very fact that $\Pi$ depends on
$\gamma$, diverging in the limit $\gamma\to 0$, while $g$ does not
require any parameter like $\gamma$ and is bounded, $g\le
1$). Therefore, the different scaling behavior of $\Pi$,
Eq.~(\ref{e2}) and $\langle g\rangle$, Eq.~(\ref{cond1a}), is not
surprising.

Returning to the SQH transition, we are thus naturally led to a
question: why do  $\Pi$ and  $\langle g\rangle$ scale identically in
this case? The reason is that the zero-energy diffusion propagator
$\Pi(e',e)$ is in fact defined at $\gamma=0$ ({\it i.e} there is no
need to introduce absorption or a finite frequency to regularize it),
see Eqs.~(\ref{e12}) and (\ref{e52}), (\ref{e54}). This can be traced
back to vanishing of DOS at $\epsilon=0$. It is not a surprise that in this
situation, when the absorption is irrelevant, $\Pi(e',e)$ and $\langle
g(e',e)\rangle$ (which only differ in the way the absorption is
incorporated) scale in the same way. 

Let us consider now higher moments 
$$
\langle[{\rm Tr} G(e',e;e^{-\gamma})G(e,e';e^\gamma)]^q\rangle.
$$ 
Applying the
consideration of Sec.~\ref{s4.3}, we find that the absorption
($\gamma$) remains irrelevant provided
\be
\label{cond4}
2qx_\rho+\Delta_{2q}>0,
\ee
with the result
\be
\label{cond5}
\langle[{\rm Tr} G(e',e;e^{-\gamma})G(e,e';e^\gamma)]^q\rangle \sim
r^{-2(qx_\rho+\Delta_q)}.
\ee
For the SQH case, the condition (\ref{cond4}) implies $q < 3/2$. (We
make an assumption that our consideration, which is strictly speaking
performed for integer $q$, remains valid for intermediate, non-integer
values of $q$.) According to the above argument, $\langle g^q\rangle$
scales in this regime in the same way (\ref{cond5}), so that 
(see also \cite{bernard01b}),
\be
X_q = 2qx_{\rho}+2\Delta_q,
\label{e16}
\ee
and, using $x_\rho=1/4$ and Eq.~(\ref{num3}), 
\be
\label{e17}
X_q \simeq q(3-q)/4.
\ee
Note that, in contrast to ${\rm Tr}GG$, the two-terminal conductance
$g$ is bounded from above, $g\le 2$ (the factor two is due to spin
summation and is not essential). Physically, it simply means that
for such rare realizations when  ${\rm Tr}GG$ is large, $g$ is limited
by the contact resistance. It follows that the exponent for $\langle
g^q\rangle$ should be a non-decreasing function of $q$. In other
words, the exponent $X_q$ saturates after reaching its maximum at some 
$q_0$. We find from (\ref{e17}) $q_0\simeq 3/2$; for larger $q$ the
exponent saturates at the value 
$X_{q\ge q_0}=X_{q_0}\simeq 9/16$ (these moments
are determined by the probability to find $g\sim 1$). 

Equation (\ref{e17}) implies, in particular, a normal distribution of
$\ln g$ at $r\gg 1$ with the average 
$\langle \ln g(r)\rangle = - X_t\ln r$ and the variance 
${\rm var}[\ln g(r)] = b \ln r$, where $X_t \simeq 3/4$ and
$b \simeq 1/2$. These values correspond to the parabolic approximation 
(\ref{num3}); more accurate predictions can be obtained by using the
numerical results for $\Delta_q$,
\bea
&& X_t = X'_0 = 2x_\rho+2\Delta'_0=2x_\rho+2(\alpha_0-2)\simeq 0.774,
\label{cond6} \\
&& b =- X''_0 = - 2 \Delta''_0 \simeq 0.58
\label{cond7}
\eea
(here a prime denotes the derivative with respect to $q$).

\begin{figure}
\centerline{
\includegraphics[width=0.85\columnwidth,clip]{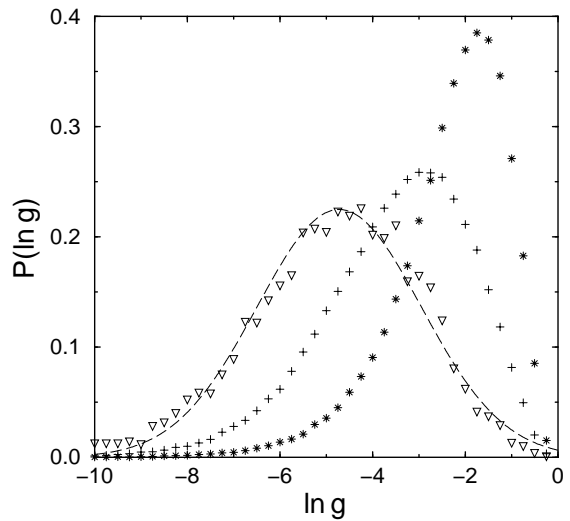}}
\vspace{3mm}
\caption{Distribution of the two point conductance at 3 different
distances between the contacts, $r=5.7 (*)$, $14.1 (+)$, $133
(\bigtriangledown)$; the system size is $L=196$. 
The dashed line indicates a log-normal fit with parameters 
$\langle \ln g\rangle=-4.72$ and ${\rm var}(\ln g)=3.15$.
}
\label{fig7}
\end{figure}

\begin{figure}
\centerline{
\includegraphics[width=0.85\columnwidth,clip]{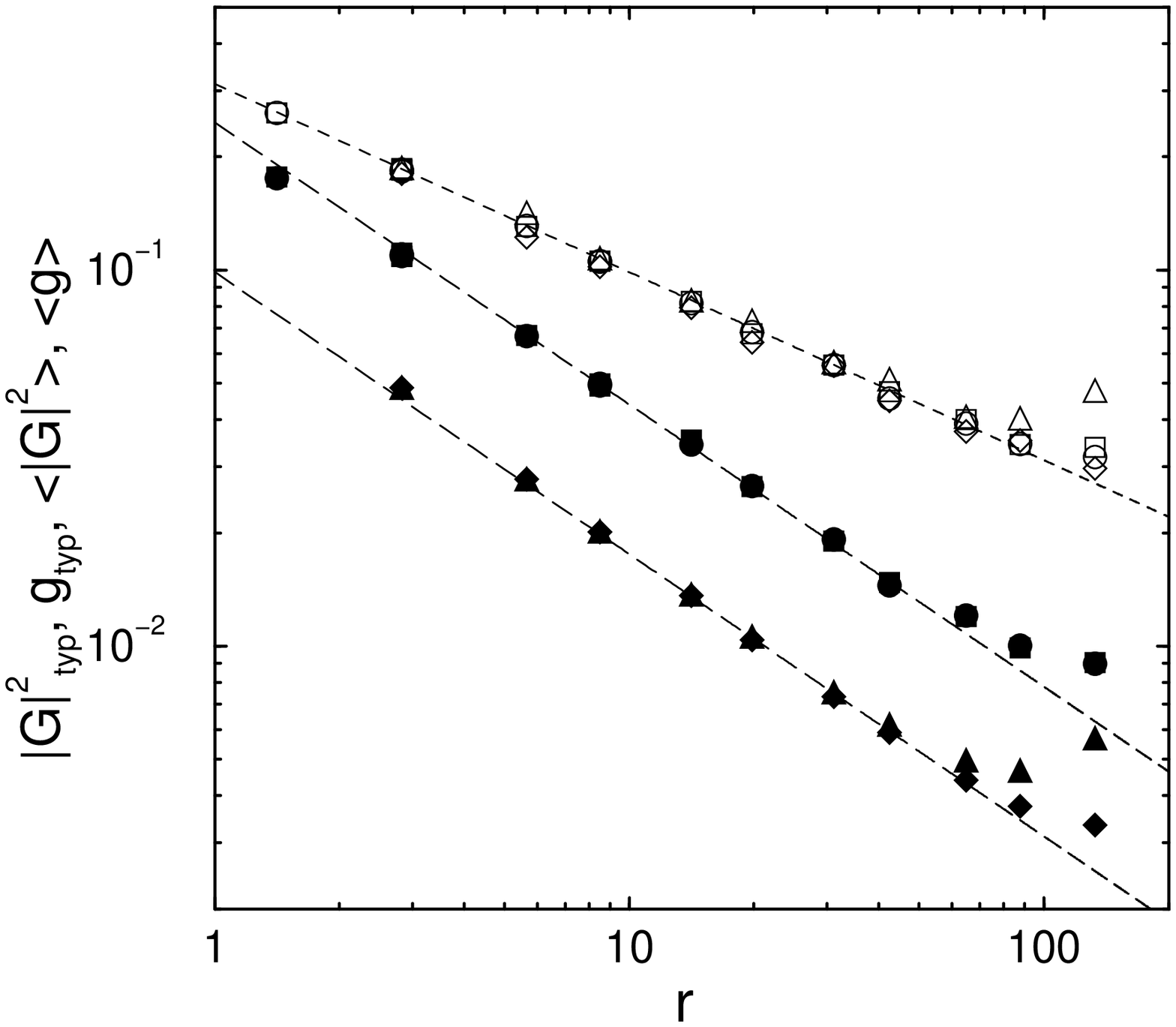}}
\vspace{3mm}
\caption{Scaling of the two-point conductance
with distance $r$ between the contacts: 
average value (empty symbols), $\langle g\rangle$,
and typical value (filled symbols),
$g_{{\rm typ}}=\exp \langle \ln g\rangle$,
in systems of sizes $L=128 (\Box)$ and $L=196 (\circ)$.
Also shown is scaling of the two-point Green function, $\langle
|G|^2 \rangle$ and 
$|G|^2_{{\rm typ}}=\exp \langle \ln |G|^2 \rangle$ 
($L=128 (\bigtriangleup), L=196 (\diamond)$). The
lines correspond to the $r^{-1/2}$ (dotted) and $r^{-3/4}$ (dashed)
power laws. Deviations from power-law
scaling at large values of $r$ are due to the finite system size. }
\label{fig8}
\end{figure}

We turn now a to a numerical study of the two-point conductance. While
we did not attempt a high-precision numerical determination of the
spectrum of corresponding exponents $X_q$ (as presented in
Sec.~\ref{s5.1} for the multifractal spectrum of wave functions), we
have verified some of the key predictions of the above analytical
considerations. Figure~\ref{fig7} illustrates evolution of the
distribution function ${\cal P}(g)$ with the distance $r$ between the
contacts; it is seen that at sufficiently large $r$ the distribution
becomes log-normal as expected. In Fig.~\ref{fig8}
we show the scaling of the average $\langle g\rangle$ and the typical
$g_{\rm typ}=\exp\langle\ln g\rangle$ values of the two-point
conductance, along with analogous quantities
$\langle |G|^2\rangle$ and 
$|G|^2_{\rm typ}=\exp\langle\ln|G|^2 \rangle$ for a closed system,
$|G|^2\equiv-{\rm Tr}G(e',e;1)G(e,e';1)$. For the average values, 
$\langle g\rangle$ and $\langle |G|^2\rangle$, the numerics fully
confirm the theoretical results (\ref{cond3}), (\ref{e9}) telling us
that the 
both quantities scale as $r^{-1/2}$ and, moreover, are equal to each
other. A non-trivial character of the equality $\langle
g\rangle=\langle |G|^2\rangle$ is well illustrated by the data for
typical quantities: $g_{\rm typ}$ and $|G|^2_{\rm typ}$ are not
equal. Nevertheless, they are found to share a common scaling: 
$g_{\rm typ}, |G|^2_{\rm typ}\sim r^{-X_t}$, confirming our arguments
presented above. Furthermore, the numerically obtained value of the
exponent, $X_t\simeq 3/4$, is in agreement with the theoretical
prediction (\ref{cond6}).

\section{Summary}
\label{s7}

Let us summarize the main results of the paper.

\begin{enumerate}

\item We have extended the mapping of the SQH network model onto the
classical percolation and calculated two- and three-point correlation
functions at the SQH transition. This allowed us to determine
analytically the fractal exponents $\Delta_2$ and $\Delta_3$ governing
the scaling of the second and third moments of the wave function
intensity, with the results $\Delta_2\equiv -\eta= -1/4$ and
$\Delta_3=-3/4$. 

\item We have performed a thorough numerical study of the multifractal
spectrum $\Delta_q$. The obtained spectrum is given with a good
accuracy by the parabolic law (\ref{num3}) but shows clear
deviations from parabolicity, Fig.~\ref{fig3}. 

\item Statistical properties of generalized inverse participation
ratios $P_q$ at the SQH transition are similar to those found earlier for
other localization transitions. In particular, the distribution
function ${\cal P}(P_q)$ becomes scale-invariant in the limit of 
large system size.  

\item We have analyzed statistics of the two-point conductance $g$ at
the localization transition with a critical density of
states. Specifically, we have presented scaling arguments which link
the exponents $X_q$ governing the spatial decay of $\langle
g^q\rangle$ to the wave-function multifractality spectrum $\Delta_q$,
see Eq.~(\ref{e16}). This yields, in particular, for the typical
conductance at the SQH critical point $g_{\rm typ}\sim r^{-X_t}$ with
$X_t\simeq 3/4$ (see Eq.~(\ref{cond6}) for a more accurate value), as
confirmed by numerical simulations. 

\end{enumerate}

In recent years, a considerable progress has been made in
understanding of conformal field theories related to problems of
two-dimensional fermions subject to quenched disorder
\cite{mudry96,zirnbauer99,
bhaseen00,fendley00,bernard01a,bernard01b,
ludwig94,nersesyan95,caux98,gurarie99,read01,altland02,gurarie02}.
In particular, a relation between the wave function multifractality in
two-dimensional 
disordered systems and the operator content of corresponding conformal 
field theories has been discussed in a number of publications
\cite{mudry96,caux98,bhaseen00,bernard01b}.
It remains an open question whether the multifractal exponents
$\Delta_q$, $X_q$ for the SQH transition can be computed by the
conformal field theory methods.  Note that our results are
against the proposal of Ref.~\cite{bernard01b}, where the result
$\Delta_q=q(1-q)/4$ was obtained. Apparently, this indicates that the theory
considered in \cite{bernard01b} and obtained \cite{bernard01a} from a
particular network model with fine-tuned couplings, does not belong
to the SQH universality class.

\section*{Acknowledgments}

Discussions and correspondence with J.T.~Chalker, I.A.~Gruzberg,
A. LeClair, A.W.W.~Ludwig, J.E.~Moore, and D.G.~Polyakov 
are gratefully acknowledged. 
This work was supported by the Schwerpunktprogramm 
"Quanten-Hall-Systeme"  and the SFB 195 der Deutschen
Forschungsgemeinschaft. 

\appendix
\section{Proofs to the mapping onto percolation}

In Sec.~\ref{s4.1} two statements were formulated which allow us to
calculate averaged products of two (Sec.~\ref{s4.1}) and three
(Sec.~\ref{s4.2}) Green functions. Here we give some more details on
the proofs of these statements. In the end of the Appendix we will
also explain why our calculation cannot be extended on products of
$q\ge 4$ Green functions (and thus on higher moments of wave functions). 

\subsection*{Statement 1}

The first statement says that only paths visiting each node 0 or 2
times are to be considered. Its proof for the case of two-point
functions was given in Sec.~\ref{s4.1}. The analysis of the case of
three-point functions (considered in Sec.~\ref{s4.2}) goes
along similar lines, and we present its brief outline only. In analogy
with (\ref{e27}), we have to consider an expression of the type
\be
\label{ea1}
\sum_{k_1,k_2,k_3=1}^\infty\langle{\rm Tr} B_1 U_f^{k_1} B_2 U_f^{k_2}
B_3 U_f^{k_3}\rangle |A(f,f)|^{k_1+k_2+k_3},
\ee
where $k_i$ is the number of returns of the $i$-th path ($i=1,2,3$) to
the edge $f$. Performing averaging over $U_f$ as in (\ref{e28}), we
cast (\ref{ea1}) into the following form
\bea
\label{ea2}
&& \sum_{k_1,k_2,k_3=1}^\infty (b_1c_{k_1+k_2+k_3}+b_2c_{k_1+k_2-k_3}
 +b_3c_{k_1+k_3-k_2}\nonumber \\ 
&& \qquad\qquad
+b_4c_{k_2+k_3-k_1})|A(f,f)|^{k_1+k_2+k_3}. 
\eea
The first term in curly brackets is trivially zero in view of
(\ref{e24}).
To demonstrate that remaining terms give zero as well, we perform a
summation over $k_i$ at fixed $k=k_1+k_2+k_3$. Indeed, it is not
difficult to show by a straightforward arithmetics that for an
arbitrary $k$
\be
\label{ea3}
\sum_{k_{1,2,3}=1,2,\ldots;\ k_1+k_2+k_3=k} c_{k_1+k_2-k_3} = 0.
\ee
Therefore, the sum (\ref{ea2}) is equal to zero, which completes the
proof of the statement 1 for the three-point Green functions.

\subsection*{Statement 2}

The second statement allows us to reduce the nodes visited 4 times to a
superposition of contributions (i) and (ii) of Fig.~\ref{fig2}, with
the factors $\cos^2\theta$ and $\sin^2\theta$, respectively. We will
give the proof for the (most non-trivial) case of a product of $q=3$
Green functions; the proof for $q=1$ and 2 is obtained in the same
way. More specifically, we will consider the correlation function
(\ref{e41}); the correlator (\ref{e40}) is treated analogously. 

\begin{figure}
\centerline{
\includegraphics[width=0.95\columnwidth]{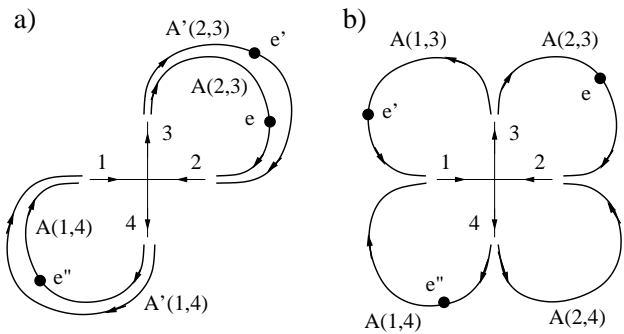}}
\vspace{3mm}
\caption{Configurations of paths for the correlation function
(\ref{e41}) for a node visited four times.}
\label{figa1}
\end{figure}

Each of three Green functions in (\ref{e41}) generates a sum over
closed loops ($e\to e$, $e'\to e'$, and $e''\to e''$,
respectively). For a given lattice node, let us label the
corresponding incoming edges as (1,2) and the outgoing ones as
(3,4). We are considering a contribution of paths visiting this node
in total four times. This generates four path segments starting each
on one of the edges (3,4) and ending on one of the edges (1,2), and not
passing through any of these edges. We are going to show that for any
configuration of these four segments the statement 2 holds. It is easy
to see that there exist two essentially different types of such
configurations (shown in Fig.~\ref{figa1}); all others can be obtained
by permutations of $e,\ e'$, and $e''$, and/or by lattice symmetry
operations.

Consider first the configuration (a) of Fig.~\ref{figa1}. We have to
sum over all ways to connect the four path segments by various
configurations of scattering events at this node shown in
Fig.~\ref{fig2}. Specifically, only such connections are allowed which
generate exactly 3 closed loops, each containing one of the edges $e$,
$e'$, and $e''$. There are three possibilities how this can be done,
since the segment that does not contain any of the edges $e,e',e''$
can be connected in a loop with any of the remaining three. In
one of these cases the configuration of paths at the considered node
is of the type (ii) of Fig.~\ref{fig2}, in two other cases it is of
the type (iii). We thus get the following contributions:
\bea
&&\langle{\rm Tr}U_2 A(2,3)U_3 \: {\rm Tr}U_2 A'(2,3)U_3 \nonumber \\
&&\qquad \times {\rm Tr} U_1A(1,4)U_4U_1A'(1,4)U_4\rangle \times s^4,
\label{ea4} \\
&&\langle{\rm Tr}U_2 A(2,3)U_3 \: {\rm Tr}U_1 A(1,4)U_4 \nonumber \\
&&\qquad \times {\rm Tr} U_2A'(2,3)U_3U_1A'(1,4)U_4 \rangle\times (-c^2s^2),
\label{ea5} \\
&&\langle{\rm Tr}U_2 A'(2,3)U_3 \: {\rm Tr}U_1 A(1,4)U_4 \nonumber \\
&&\qquad \times {\rm Tr} U_2A(2,3)U_3U_1A'(1,4)U_4 \rangle\times (-c^2s^2).
\label{ea6} 
\eea
Here $A(2,3)$ is a sum over all paths from 3 to 2 passing through $e$, 
$A'(2,3)$ is a sum over paths $3\to e'\to 2$, 
$A(1,4)$ is a sum over paths $4\to e''\to 1$, and  
$A'(1,4)$ is a sum over paths $4\to 1$ (Fig.~\ref{figa1}a). Also, we
have denoted $s=\sin\theta$ and $c=\cos\theta$.

To perform the integration over $U_i$, 
we use the following formulas of integration over SU(2) matrices:
\bea
\langle {\rm Tr} UV_1 \: {\rm Tr} UV_2\rangle_U & = &
{1\over 2}{\rm Tr}V_1^\dagger V_2, \label{ea7} \\
\langle {\rm Tr} UV_1 UV_2\rangle_U & = &
-{1\over 2}{\rm Tr}V_1^\dagger V_2, \label{ea8} \\
\langle {\rm Tr} UV_1 U^\dagger V_2\rangle_U & = &
{1\over 2}{\rm Tr}V_1 \: {\rm Tr} V_2. \label{ea9} 
\eea
Here matrices $V_{1,2}$ are assumed to be of the form
$V_i=|V_i|\tilde{V_i}$, where $\tilde{V_i}\in SU(2)$ and $|V_i|$ is a
real number (we remind that $A(i,j)$ are exactly of this type, see
Sec.\ref{s4.1}). 

Applying repeatedly the rules (\ref{ea7})--(\ref{ea9}), we perform
integration over all matrices $U_i$ ($i=1,\ldots,4$) in
Eqs.~(\ref{ea4})--(\ref{ea6}). We find then that all three
contributions  (\ref{ea4})--(\ref{ea6}) are proportional to 
${\rm Tr} A^\dagger(2,3)A'(2,3)\: {\rm Tr} A^\dagger(1,4)A'(1,4)$, with
coefficients $-{1\over 4}s^4$, $-{1\over 8}c^2s^2$, and  
$-{1\over 8}c^2s^2$, respectively. The total coefficient is therefore
\be
\label{ea10}
-{1\over 4}s^4-{1\over 8}c^2s^2-{1\over 8}c^2s^2 = -{1\over 4}s^2.
\ee
We see that the same result would be obtained if we would assign the
weight $s^2$ to the first contribution (which is of the type (ii) of
Fig.~\ref{fig2}) and discard the remaining two terms (which are of the
type (iii)). This establishes the validity of the statement 2 with
respect to the configuration (a) of Fig.~\ref{figa1}.

The configuration (b) of Fig.~\ref{figa1} is analyzed along the same
lines. We have again three contributions, one of the type (ii) of
Fig.~\ref{fig2} and two of the type (iii),
\bea
&&\langle{\rm Tr}U_2 A(2,3)U_3 \: {\rm Tr}U_1 A(1,4)U_4 \nonumber \\
&&\qquad \times {\rm Tr} U_1A(1,3)U_3U_2A(2,4)U_4\rangle \times s^4,
\label{ea11} \\
&&\langle{\rm Tr}U_2 A(2,3)U_3 \: {\rm Tr}U_1 A(1,3)U_3 \nonumber \\
&&\qquad \times {\rm Tr} U_1A(1,4)U_4U_2A(2,4)U_4 \rangle\times (-c^2s^2),
\label{ea12} \\
&&\langle{\rm Tr}U_1 A(1,3)U_3 \: {\rm Tr}U_1 A(1,4)U_4 \nonumber \\
&&\qquad \times {\rm Tr} U_2A(2,3)U_3U_2A(2,4)U_4 \rangle\times (-c^2s^2).
\label{ea13} 
\eea
After integration over $U_i$ according to the rules
(\ref{ea7})--(\ref{ea9}) they all produce an identical structure,
${\rm Tr} A^\dagger(2,3)A(2,4)A^\dagger(1,4)A(1,3)$, with the
coefficients ${1\over 4}s^4$, ${1\over 8}c^2s^2$, and  
${1\over 8}c^2s^2$, respectively. Again, retaining only the (ii)-type
contribution (\ref{ea11}) and assigning the weight $s^2$ to it, we
would obtain the same result. This completes the proof of statement 2
for the three-point correlation function (\ref{e41}). 

\subsection*{What about $q>3$?}

A natural question is whether the present approach can be generalized 
to higher-order correlations of wave functions governed by 
multifractal exponents $\Delta_q$ with $q>3$. The answer is
negative. In fact, both statements 1 and 2 do not apply 
(or, in a more careful formulation, our proofs fail) for $q\ge 4$, as
we are going to explain in brief. Concerning the statement 1, consider
a generalization of the expression (\ref{ea1}) to $q=4$, and choose $k\equiv
k_1+k_2+k_3+k_4=4$. Obviously, there is just one such term (all $k_i=1$) in
the sum, and it is easy to see that it is generically
non-zero. Therefore, no cancellation of terms with $k>2$ happens in
this case, {\it i.e.} the statement 1 does not work. 
Turning to the statement 2, consider {\it e.g.} a correlation function
$\tilde{D}(e,e',e'',e''';\gamma)$ analogous to (\ref{e41}) but
containing a product of four traces of Green functions. Trying to
prove the statement 2, we will then have to consider the path
configurations very similar to those shown in Fig.~\ref{figa1} but
with all four paths containing one of the edges $e$, $e'$, $e''$, or
$e'''$. At the next step the paths should be connected via the
scattering processes at the node -- this time to generate 4 closed
loops. However, for each of the configurations shown in
Fig.~\ref{figa1} there is only one way to do this, so that only one
contribution will arise in place of three terms (\ref{ea4})--(\ref{ea6})
or (\ref{ea11})--(\ref{ea13}). Clearly, the statement 2 is not valid in
this situation. Therefore, the mapping onto the classical percolation
is not applicable for higher moments, $q>3$. This is in correspondence
with the fact that $q=3$ separates two regimes of qualitatively
different behavior of correlation functions, as discussed in
Sec.~\ref{s4.3}.

\end{multicols}
\end{document}